\documentclass[iopart12]{iopart}
\usepackage{iopams}  
 \usepackage{xcolor}
 \usepackage{graphicx}
 \usepackage{lipsum}

\usepackage{array}   
\newcolumntype{C}{>{$}c<{$}} 
\newcolumntype{L}{>{$}l<{$}} 
\usepackage{hhline}
\usepackage{caption}
\usepackage{mathrsfs}
\DeclareMathAlphabet{\mathpzc}{OT1}{pzc}{m}{it}
\usepackage{amssymb}
\usepackage{float}
\usepackage{bbm,dsfont}
\usepackage{graphicx}
\usepackage{amsfonts}
\usepackage{graphicx}  
\usepackage{color}
\usepackage{soul}
\setstcolor{blue}
\usepackage{todonotes}

\def\wk{\hat{\mathpzc{w}}_\lambda(\Omega)}

\def\wkzero{\hat{\mathpzc{w}}_\lambda(0)}
\def\wz{\hat{\mathds{S}}}

\def\wh{\hat{\mathds{S}}^{(2)}}
\def\ac{\mathpzc{a}}

\def\Cop{\hat{\mathpzc{C}}\color{black}}

\newcommand{\hdg}[1]{\textcolor{red}{#1}}

\def\beq{\begin{equation}}
\def\eeq{\end{equation}}
\def\beqa{\begin{eqnarray}}
\def\eeqa{\end{eqnarray}}
\def\hf{\textstyle{1\over2}}
\def\3hf{\textstyle{\frac{3}{2}}}
\def\half{{1\over2}}

\newcommand{\cg}[6]{\mbox{$\Bigl\langle {#1 \atop #2} ; {#3 \atop #4} \,\big\vert\, {#5 \atop #6} \Bigr\rangle$}}
\newcommand{\rcg}[6]{\mbox{$\Bigl\langle{#1 \atop #2} ;{ #3 \atop #4} \, \Vert\, {#5 \atop #6}\Bigr\rangle $}}
\newcommand{\sixj}[6]{\mbox{$\left\{\begin{array}{ccc} #1 & #2 & #3\\ #4 & #5 & #6 \end{array}\right\}$}}

\newcommand{\ket}[1]{\vert #1 \rangle}
\newcommand{\bra}[1]{\langle #1 \vert}
\renewcommand{\e}{\hbox{\rm e}}

\def\hdg#1{\color{red}  #1 \color{black} }
\def\andrei#1{\color{red} #1 \color{black}}
\def\hdg#1{\color{black}  #1 \color{black} }
\def\andrei#1{\color{black} #1 \color{black}}

\begin{document}

\title{Correspondence rules for Wigner functions over $SU(3)/U(2)$}
\author{Alex Cl\'{e}sio Nunes Martins$^\dagger$, Andrei B. Klimov$^\ddagger$, and Hubert 
de Guise$^\dagger$}

\address{$^\dagger$ Department of Physics, Lakehead University, Thunder
Bay, Ontario P7B 5E1, Canada \\
$^\ddagger$ Departamento de F\'{i}sica, Universidad de Guadalajara, 44420 Guadalajara, Mexico}

\begin{center}
\today 
\end{center}

\begin{abstract}
We present results on the $\star$ product for $SU(3)$ Wigner functions over $SU(3)/U(2)$.  In particular, 
we present a form of the so-called correspondence rules, which provide a differential form of the 
$\star$ product $\hat A\star \hat B$ and $\hat B\star \hat A$ when $\hat A$ is an $\mathfrak{su}(3)$ generator.  
\hdg{For the $\mathfrak{su}(3)$ Wigner map, 
these rules must contain second order derivatives and thus substantially differ from the rules
of other known cases.} 
\end{abstract}

\date{September 2018}



\section{Introduction}

The possibility {of ``switching on''} our classical intuition in the analysis of 
quantum systems is an important advantage of the phase space approach to quantum mechanics \cite{revPS}. 
Given this, the phase-space approach {has been} directly and efficiently applied to 
``large'' (semi-classical) quantum systems with Heisenberg-Weyl $HW(n)$, 
rotation $SO(3)$ or Euclidean $E(2)$ dynamical symmetries. The corresponding classical 
manifolds ($2n$-dimensional plane, $S^2$ sphere and 2-dimensional cylinder) are easy to visualize, and
quantum states $\hat \rho$ can be conveniently mapped to corresponding 
(quasi-) distribution functions $W_\rho(\Omega)$, where $\Omega$ is a phase-space 
point. 

\andrei{For quantum systems with observables in the algebra of the 
$SU(n)$ group acting in a Hilbert space that carries a symmetric unitary
irreducible representation $(\lambda ,0...,0)$, there is a systematic
construction \cite{suNKdeG} of phase-space functions satisfying the basic
Stratanovich-Weyl requirements \cite{WF,brif,Polkovnikov} (see also \cite{Tilma}).}

In this paper, we concentrate on the Wigner representation \cite{WF}\cite{brif}\cite{Polkovnikov}, where an 
operator $\hat f \mapsto W_{\hat f}(\Omega)$ is mapped \andrei{into a self-dual symbol $W_{\hat{f}}(\Omega )$,%
\begin{equation}
\hat{f}\Leftrightarrow W_{\hat{f}}(\Omega ),\quad \Omega \in \mathfrak{M,}
\label{map}
\end{equation}%
where $\mathfrak{M}=SU(n)/U(n-1)$ is a symplectic manifold corresponding to
a classical phase space \cite{Onofri}, intimately related with the set of
orbit-type coherent states \cite{coherentstatestuff} generated from a highest weight state 
of the corresponding Hilbert space.}   
Of the several types of existing  phase-space maps, the Wigner representation is of particular interest 
as not only sensitive to the interference pattern but also endowed with 
certain important dynamical properties \cite{revPS}\cite{WF}\cite{book}\cite{review}.

\andrei{The map of Eq.(\ref{map}) has a relatively simple form for the fundamental representation of $SU(n)$, but its explicit
construction for arbitrary symmetric representations of $SU(n)$ requires detailed knowledge of the appropriate Clebsch-Gordan
technology and is more involved \cite{suNKdeG,review,CGAlex}. In spite of this, the limit of large dimensions where 
$\lambda\gg 1$ is of immediate interest to this work as it is related to the semi-classical description of quantum systems, and
is especially important for the analysis of the evolution of macroscopic systems since $\lambda$ is often identified with the number
of particle constituents in the system.}   
\andrei{ Moreover, whereas the dimension of the symmetric irrep of $SU(n)$ grows with $\lambda$ like $\lambda^{n-1}$, 
the dimension of the classical phase-space $SU(n)/U(n-1)$ is $2(n-1)$ and thus independent of $\lambda$ and 
grows linearly with $n$; phase-space methods thus become increasingly efficient as $\lambda$ increases.  In the 
limit $\lambda\gg 1$, the evolution equation for the Wigner distribution admits a natural expansion in a single
semi-classical parameter $\epsilon$, which scales as $\lambda^{-(n-1)/2}$, as can be seen for $SU(3)$ in Eq.(\ref{SP}).}
\andrei{This classical limit} \hdg{of large dimensions} \andrei{is well-studied in
the case of $SU(2)$, describing} \hdg{very pictorially}  \andrei{spin-like systems \cite{review}}
\hdg{as quasi-probability distributions on the sphere $S^2$}, \andrei{
but the situation is far from transparent for higher-rank unitary groups.} 

{The} phase-space picture is 
complete when a sensible expression for the star-product  $\star $, 
defining the composition map $\hat f \hat g\mapsto W_{\hat f}(\Omega) 
\star W_{\hat g}(\Omega)$, is found \cite{Moyal}\cite{Belchev}\cite{su2}.
The \hdg{(non-commutative)} $\star$-product operation \hdg{is essential to capture the non-commutative nature of quantum 
mechanical operators and} can be used to map 
the Schr\"{o}dinger equation into a Liouville-type 
equation of motion for the Wigner function $W_{\hat \rho}(\Omega)$ \cite{TWA}. Our experience with  $SU(2)$ systems \cite{klimov}
suggests that   {this Liouville-type equation} 
can be efficiently expanded {in powers of} \hdg{the} semiclassical 
parameter, even for 
{systems with $SU(n)$ symmetry.}
In this expansion, the leading term is {a} first-order differential 
operator describing the classical dynamics of the Wigner distribution and the 
first-order corrections to the classical motion vanish. 
For Hamiltonians polynomial in the generators $\hat C_{\alpha}$ of the corresponding  {$\mathfrak{su}(n)$} Lie algebra 
the star-product can be 
{replaced} by the {more specialized}  
correspondence rules (or Bopp operators) \cite{boop}, that 
establish particular maps
\begin{equation}
\hat C_{\alpha} \hat f \rightarrow 
\hat \mathfrak{C}_{\alpha}^{L}W_{\hat f}(\Omega) ,\qquad 
\hat f \hat C_{\alpha} \rightarrow \hat\mathfrak{C}_{\alpha}^{R} W_{\hat f}(\Omega)
\end{equation}
where $\hat\mathfrak{C}_{\alpha}^{L}$ and $\hat\mathfrak{C}_{\alpha}^R$ are some differential operators. 

In the $SU(2)$ case \cite{review}\cite{stratonovich}\cite{bondia} 
the operators $\hat\mathfrak{C}_{\alpha}^{L,R} $ have the form of first-order differential operators 
{multiplied by} functions of the Casimir operators \cite{book}\cite{su2CR}.  
{This} form of $\hat\mathfrak{C}_{\alpha}^{L,R}$ operators is related to the 
{simple} structure of the  
classical phase-space isomorphic to the coset $SU(2)/U(1)$ and indeed
closely connected to $SU(2)$ coherent states \cite{coherentstatestuff}.

For systems with higher unitary symmetries, the explicit differential realizations of the $\star$-product and of the operators
$\hat\mathfrak{C}_{\alpha}^{L,R}$ is still an open question.
{Indeed for systems with symmetries beyond $SU(2)$, $E(2)$ or $HW(n)$,} 
the situation with phase-space mapping becomes significantly more 
abstract \cite{suNKdeG} \cite{suN}. Even in the simple case 
of a system transforming by a symmetric irreducible representation $(\lambda,0)$ of the $SU(3)$ group, the Wigner 
map to the symplectic manifold $\mathfrak{M} = SU(3)/U(2)$ is quite involved \cite{suNKdeG}. 
Some relevant information from four-dimensional quasi-distribution functions can be extracted by 
appropriately projecting them into two-dimensional submanifolds, as was done in \cite{su3evol}.  Nevertheless it is remarkable that 
a single semi-classical parameter, 
{easily related to the square root of the eigenvalue of quadratic Casimir operator of $\mathfrak{su}(3)$ acting in the}
representation $(\lambda,0)$, still appears very naturally in the  Wigner map \cite{review}.
      
In this paper we obtain the correspondence rules for the $SU(3)$ Wigner function defined on the phase
space $SU(3)/U(2)$, appropriate to the most physically relevant  case of the symmetric representation 
$(\lambda,0)$ of $SU(3)$.
We find that, contrary to the $SU(2)$ and other known cases, the correspondence rules require second order derivatives 
(in addition to the differential action of the $\mathfrak{su}(3)$ Casimir operator),
highlighting a fundamental difference between the present work and all previous results \cite{WF}.  We ascribe this difference to
the representation theory of the $U(2)$ subgroup that leaves invariant the highest weights of
$SU(3)$ irreps of the type $(\lambda,0)$; this subgroup
can accommodate representations of dimension greater than $1$.  We apply 
{the correspondence rules to}
the equations of motion for some non-linear Hamiltonians {quadratic in the 
$\mathfrak{su}(3)$ generators.} 
We find the form of the correspondence rules in the limit of large dimension of 
the representations, which allows us to obtain the semiclassical expansion of the equations 
of motion for Hamiltonian quadratic and more generally polynomial in the $\mathfrak{su}(3)$ generators.
 
\section{The $SU(3)$ Wigner function}

The group $SU(3)$ is defined as the set of unitary $3\times 3$ matrices with determinant $1$.  A convenient way
to parametrize this set is by using a slight variation of the result given in \cite{factorization} (see also \cite{generalfactorization}) 
and write an element $g\in SU(3)$ as a sequence of block $SU(2)$ matrices 
\cite{review}:
\beq 
g=\hat R_{23}(\alpha_1,\beta_1,-\alpha_1)
\hat R_{12}(\alpha_2,\beta_2,-\alpha_2)
\left[ \hat R_{23}(\alpha_3,\beta_3,-\alpha_3)
\Phi(\gamma_1,\gamma_2)\right]
\eeq
where $\Phi$ is a diagonal matrix with determinant $1$ containing two independent phases. Note that the
last bracketed factors 
\beq 
\left[ \hat R_{23}(\alpha_3,\beta_3,-\alpha_3)
\Phi(\gamma_1,\gamma_2)\right] \label{U2subgroup}
\eeq 
are elements of an $U(2)\sim SU(2)\otimes U(1)$ subgroup.
The corresponding 
 $\mathfrak{su}(3)$ Lie algebra is spanned by the eight generators
\beqa 
\hat C_{12},\hat C_{23},\hat C_{13}\, , &&\quad  \hbox{3 raising operators}\, ,\nonumber \\
\hat h_1=\hat C_{22}-\hat C_{33}\, ,\hat h_2=2\hat C_{11}-\hat C_{22}-\hat C_{33}\, ,&&
\quad \hbox{2 Cartan generators}\, ,\label{eq:su3generators} \\
\hat C_{21},\hat C_{32},\hat C_{31}\, ,&&\quad  \hbox{3 lowering operators}\, , \nonumber 
\eeqa
with the basic commutation relations
\beq 
\left[\hat C_{ij},\hat C_{k\ell}\right]=\delta_{jk}\hat C_{i\ell}- \delta_{i\ell} \hat C_{kj}
\eeq 

With this we can recall some 
\hdg{elementary facts on} the construction of the symmetric $SU(3)$ Wigner function for $SU(3)$ irreps of the 
type $(\lambda,0)$ \cite{suNKdeG} \cite{factorization}.  Basis states in the corresponding $\frac{1}{2}
(\lambda+1)(\lambda+2)$ dimensional Hilbert space
\beq
\ket{(\lambda,0)\nu I}:=\ket{(\lambda,0)\nu_1\nu_2\nu_3;I} \quad , 
\eeq
are labeled by  
the triple $\nu=(\nu_1,\nu_2,\nu_3)$ of 
nonnegative occupation numbers, subject to the constraints  $\nu_1+\nu_2+\nu_3=\lambda$. 
The weight of the state  {with occupation numbers $(\nu_1,\nu_2,\nu_3)$ is $[\nu_1-\nu_2,\nu_2-\nu_3]$. 
For a general $(p,q)$ irrep, a given weight can occur more than once; in such cases an $\mathfrak{su}(2)$ label $I$
is enough to uniquely identify a set of basis states with identical weights.
For states in $(\lambda,0)$  $I$ is completely determined by the occupation numbers: $I=\frac{1}{2}(\nu_2+\nu_3)$;
as a result this  $I$ index can sometimes be conveniently omitted for states in $(\lambda,0)$.}

The  {appropriate} phase-space is isomorphic to the coset 
$SU(3)/U(2)\sim \mathbb{CP}^2$ , where the displacement operator	
\beq 
\hat R(\Omega)=\hat{R}_{23}(\alpha_1,\beta_1,-\alpha_1)\hat{R}_{12}(\alpha_2,\beta_2,-\alpha_2)\hat{R}_{23}(\alpha_1,-\beta_1,-\alpha_1), 
\label{R} 
\eeq
depends on $\Omega \in SU(3)/U(2)$, with angles in the ranges
\beq 
0\leq \alpha_{1,2}\leq 2\pi, \quad 0\leq \beta_{1,2}\leq \pi,
\eeq 
and where $\hat R_{ij}$ is an $SU(2)$ subgroup rotations, generated by 
$\{\hat C_{ij},\hat C_{ji}, [\hat C_{ij},\hat C_{ji}]\}$.

{If states transform by the irrep $(\lambda,0)$, operators acting on these states
will be of the generic form $\ket{(\lambda,0) \nu;I}\bra{(\lambda,0)\nu';I'}$ 
and will transform by the irrep $(\lambda,0)\otimes (0,\lambda)$, where $(0,\lambda)$ is the irrep
conjugate to $(\lambda,0)$. This representation is reducible and decomposes in the direct sum \cite{su3decomposition}
\beq 
(\lambda,0)\otimes (0,\lambda)=\sum_{\sigma=0}^{\lambda}(\sigma,\sigma)\, ,
\eeq
with $(\sigma,\sigma)$ irreducible.}
The  Wigner symbol of an arbitrary operator $\hat A$ acting on $\ket{\lambda;\nu}$ 
states is then defined as  
\beq 
W_{\hat A}(\Omega)=\hbox{Tr}(\hat{\mathpzc{w}}_\lambda(\Omega) \hat A)
\label{WF}
\eeq 
where  the kernel $\hat{\mathpzc{w}}_\lambda(\Omega)$ has the form
\beqa 
\hat{\mathpzc{w}}_\lambda(\Omega)&=\sum_{\sigma=0}^\lambda F^\lambda_{\sigma}\sum_{\nu J}D^{(\sigma,\sigma)}_{\nu J;(\sigma\sigma\sigma)0}(\Omega)\hat{T}^{\lambda}_{\sigma;\nu J}, \label{kernelomegalambda}  \\
F^\lambda_\sigma&=\sqrt{\frac{2(\sigma+1)^3}{(\lambda+1)(\lambda+2)}} \, ,
\eeqa 
with
\beqa
\hat T^{\lambda}_{\sigma;\gamma I_\gamma}&=\sum_{\nu \beta}\ket{(\lambda,0)\nu;I_{\nu}}\bra{(\lambda,0)\beta ;I_{\beta}} 
\tilde{C}^{\sigma \gamma I_\gamma}_{\lambda \nu I_\nu;\lambda^* \beta^* I_{\beta}}
\label{deftensorope} \, ,\\
I_{\nu}&= \textstyle\frac{1}{2}(\lambda-\nu_1)
\eeqa 
a component of the irreducible tensor operator transforming by $(\sigma,\sigma)$, 
and where coset (harmonic) functions, invariant under the $U(2)$ subgroup of Eq.(\ref{U2subgroup}),
 are obtained as matrix elements of (\ref{R}):
\beq 
D^{(\sigma,\sigma)}_{\nu I;%
(\sigma\sigma\sigma)0}(\Omega)\equiv \langle (\sigma,\sigma)%
\nu I \vert \hat R(\Omega)\vert (\sigma,\sigma)\sigma\sigma\sigma;0 \rangle,
\label{HF}
\eeq
The coefficients $\tilde{C}^{\sigma \gamma I_\gamma}_{\lambda \alpha I_\alpha;\lambda^* \beta^* I_{\beta^*}}$ are {up to a phase, $\mathfrak{su}(3)$ Clebsch-
Gordan coefficients;  their expressions are} given in \ref{sec:tensoroperators}.
In particular, the symbol of a tensor operator  $\hat{T}^\lambda_{\sigma;\mu J}$  is
\beq 
W_{\hat{T}^\lambda_{\sigma;\mu J}}(\Omega)=
F^\lambda_\sigma \left(D^{(\sigma,\sigma)}_{\mu  J;(\sigma\sigma\sigma)0}(\Omega)\right)^*.
\label{symT}
\eeq 
In Table \ref{table:wigsymbols} we give symbols of $\mathfrak{su}(3)$ generators of Eq.(\ref{eq:su3generators})
and their relations with the 
{$(1,1)$} tensor operators, where
\begin{equation}
N=\sqrt{\lambda(\lambda+1)(\lambda+2)(\lambda+3)}
\label{N}
\end{equation}
is a normalization factor. 

\begin{table}[H] \centering
\caption{Wigner symbols of the $su(3)$ generators}
\label{table:wigsymbols}
{\renewcommand{\arraystretch}{2.05}
\begin{tabular}{| C | C | C |}
\hline
\hat{C}_{12}&-\frac{N}{2\sqrt{6}}\hat{T}^\lambda_{1;(201)\frac{1}{2}}& \hf \sqrt{\lambda(\lambda+3)}\e^{i\alpha_2}\cos(\hf\beta_1)\sin\beta_2 \\
\hline
\hat{C}_{13}& \frac{N}{2\sqrt{6}}\hat{T}^\lambda_{1;(210)\frac{1}{2}} &\hf\sqrt{\lambda(\lambda+3)}\e^{i(\alpha_1+\alpha_2)}\sin(\hf\beta_1)\sin\beta_2 \\
\hline
\hat{C}_{23}& \frac{N}{2\sqrt{6}}\hat{T}^\lambda_{1;(120)1}&\frac{1}{2}\sqrt{\lambda(\lambda+3)}\e^{i\alpha_1}\sin\beta_1\sin^2(\hf\beta_2) \\
\hline
\hat{C}_{21} & -\frac{N}{2\sqrt{6}}\Big(\hat{T}^\lambda_{1;(201)\frac{1}{2}}\Big)^\dagger&\frac{1}{2}\sqrt{\lambda(\lambda+3)}\e^{-i\alpha_2}\cos(\frac{1}{2}\beta_1)\sin\beta_2 \\
\hline
\hat{C}_{31} & \frac{N}{2\sqrt{6}}\Big(\hat{T}^\lambda_{1;(210)\frac{1}{2}}\Big)^\dagger & \frac{1}{2}\sqrt{\lambda(\lambda+3)}\e^{-i(\alpha_1+\alpha_2)}\sin(\frac{1}{2}\beta_1)\sin\beta_2\\
\hline
\hat{C}_{32} & \frac{N}{2\sqrt{6}}\Big(\hat{T}^\lambda_{1;(120)1}\Big)^\dagger & \frac{1}{2}\sqrt{\lambda(\lambda+3)}\e^{-i\alpha_1}\sin\beta_1\sin^2(\frac{1}{2}\beta_2) \\
\hline
\hat{H}_2 =
\frac{1}{\sqrt{6}}\hat h_2& \frac{N}{2}\hat{T}^\lambda_{1;(111)0} &\frac{1}{2}\sqrt{\lambda(\lambda+3)}(1+3\cos\beta_2) \\
\hline
\hat{H}_1 
=-\frac{1}{\sqrt{2}}
\hat h_1& -\frac{N}{4\sqrt{3}}\hat{T}^\lambda_{1;(111)1} &\sqrt{\lambda(\lambda+3)}\cos\beta_1\sin^2(\frac{1}{2}\beta_2) \\
\hline
\end{tabular}
}
\label{wigCsymbols}
\end{table}

Using the transformation property of the tensor operators 
\beq 
\hat{R}(\Omega)\hat{T}^{\lambda}_{\sigma;\mu I}\hat{R}^{\dagger}(\Omega)=
\sum_{\nu J}D^{(\sigma,\sigma)}_{\nu J;\mu I}(\Omega)\hat{T}^{\lambda}_{\sigma;\nu J}
\label{transformtensors}
\eeq 
where 
\beq 
D^{(\tau,\tau)}_{\nu J;\mu I}(\tilde \Omega)=\langle (\sigma,\sigma)
\nu I \vert R(\Omega)\vert (\sigma,\sigma) \mu J  \rangle
\label{SU3D}
\end{equation}
are $SU(3)$ $D$-functions, we can represent the quantization kernel $\hat{\mathpzc{w}}_\lambda(\tilde{\Omega})$ in an explicitly covariant form
\beq 
\hat{\mathpzc{w}}_\lambda( \Omega)=\hat{R}(\Omega)\hat{\mathpzc{w}}_\lambda(0)\hat{R}^\dagger(\Omega)
\label{kernelomega}
\eeq 
{where $\hat R$ is given in Eq.(\ref{R}) and} $\hat{\mathpzc{w}}_\lambda(0)$ 
contains only  diagonal tensor operators
\begin{equation}
\hat{\mathpzc{w}}_\lambda(0)=\sum_{\sigma=0}^\lambda  F^\lambda_\sigma\hat{T}^{\lambda}_{\sigma;(\sigma\sigma\sigma)0 } .
\label{katzero}
\end{equation}
{and is invariant under $U(2)$ transformations of the type given in Eq.(\ref{U2subgroup})}.

\section{The Correspondence Rules}\label{sec:rules}

The correspondence rules allow us to represent the symbol of a product of a $\mathfrak{su}(3)$ generator 
$\hat C_{ij}$ of  Eq. (\ref{eq:su3generators}) with an arbitrary operator $\hat B$ in form of a local 
action, {\it v.g.}
\beqa
W_{\hat C_{ij} \hat B} (\Omega)&=\hbox{Tr}\left( \wk \hat{C}_{ij} \hat{B}\right)=\hat\mathfrak{C}_{ij} W_{\hat B} (\Omega),
\eeqa
where $\hat\mathfrak{C}_{ij}$ is a differential operator.  It is clear that it suffices to express the right 
and left products of first-rank tensor operators on the Wigner kernel as an operator acting on the argument of the kernel. 

Let us start with the left action. It is convenient to separate the calculations into two steps. First, we 
reduce 
the problem of arbitrary tensors multiplication to {one involving} a diagonal 
form by using Eq.(\ref{kernelomega}):
\beqa
\hat{T}^{\lambda}_{1;\alpha J}\wk  &=
\hat{T}^{\lambda}_{1;\alpha J}\hat{R}(\Omega)\wkzero    \hat{R}^\dagger(\Omega)\, \nonumber \\
&=\hat{R}(\Omega)\left[\hat{R}^\dagger(\Omega)\hat{T}^{\lambda}_{1;\alpha J}\hat{R}(\Omega)\right]\wkzero    
\hat{R}^\dagger(\Omega)\, \nonumber \\
&=\sum_{\nu I}D^{(1,1)}_{\nu I;\alpha J}(\Omega^{-1})
\left[\hat{R}(\Omega)\, \hat{T}^{\lambda}_{1;\nu I} \wkzero    \,\hat{R}^\dagger(\Omega)\right]\, .
\label{initialactleftw}
\eeqa
The possible values of $\nu I$ are given as part of Table \ref{table:tableofnunubar}.
The coupling  $\hat{T}^{\lambda}_{1;\nu I} \wkzero$ has a form of a linear combination 
\beqa
\hat{T}^{\lambda}_{1;\nu I} \wkzero    &=& \sum_{\sigma=0}^\lambda  F^\lambda_\sigma\,
\hat{T}^{\lambda}_{1;\nu I}\hat{T}^{\lambda}_{\sigma;(\sigma\sigma\sigma)0 } \\
&=& \sum_{\tau}\, F^{\lambda}_\tau\,
\ac^L_{\nu_1 I}(\lambda;\tau ) \hat{T}^{\lambda}_{\tau;\bar\nu I}
\label{actleftw}
\eeqa
where the coefficients $\ac^L_{\nu_1 I}(\lambda; \tau) $ are  
{evaluated explicitly} in 
\ref{acoefficients} and  $\nu$ and $\bar\nu$ are related in Table \ref{table:nunubar}. 

\begin{table}[H]
  \begin{center}
 \caption{Relationship between $(\nu_1\nu_2\nu_3)I$ and $(\bar\nu_1\bar\nu_2\bar\nu_3)I$.}
 \label{table:nunubar}
 {\renewcommand{\arraystretch}{1.5}
    \begin{tabular}{|c|c||c|c|}
    \hline
$(\nu_1\nu_2\nu_3) I$&$(\bar\nu_1\bar\nu_2\bar\nu_3)I$&$(\nu_1\nu_2\nu_3)I$&$(\bar\nu_1\bar\nu_2\bar\nu_3)I$ \\
    \hline
    $(210)\frac{1}{2}$&$(\tau+1,\tau,\tau-1)\frac{1}{2}$&$(111)1$&$(\tau,\tau,\tau)1$ \\
     $(201)\frac{1}{2}$ & $(\tau+1,\tau-1,\tau)\frac{1}{2}$& $(102)1$ & $(\tau,\tau-1,\tau+1)1$ \\
    $(021)\frac{1}{2}$&$(\tau-1,\tau+1,\tau)\frac{1}{2}$&$(120)1$ &$(\tau,\tau+1,\tau-1)1$ \\
    $(012)\frac{1}{2}$&$(\tau-1,\tau,\tau+1)\frac{1}{2}$&$(111)0$&$(\tau,\tau,\tau)0$ \\
    \hline
    \end{tabular}\label{table:tableofnunubar}
    }
  \end{center}
\end{table}
Thus, Eq. (\ref{initialactleftw}) is transformed   to
\beq
\kern -2em \hat{T}^{\lambda}_{1;\alpha J} \wk =   \sum_{\nu I} D^{(1,1)}_{\nu I;\alpha J}(\Omega^{-1})
\sum_{\tau \nu' I'} F^\lambda_\tau \ac^L_{\nu_1 I}(\lambda;\tau) D^{(\tau,\tau)}_{\nu'I',\bar\nu I}(\Omega) 
\hat{T}^{\lambda}_{\tau;\nu' I'}.
\label{partialactionfromtheleft}
\eeq  
 A more useful explicit form is given by 
\beqa
&\hat{T}^{\lambda}_{1;\alpha J} \wk \nonumber \\
&=\sum_{\nu_1=0,2} \sum_{\nu_2\nu_3}
D^{(1,1)}_{\nu \frac{1}{2};\alpha J}(\Omega^{-1}) \sum_{\tau \nu' I'} F^{\lambda}_{\tau}  
\ac^L_{\nu_1 \frac{1}{2}}(\lambda;\tau)
D^{(\tau,\tau)}_{\nu' I';\bar\nu \frac{1}{2}}(\Omega) \hat{T}^{\lambda}_{\tau;\nu' I'}  \nonumber \\
&\quad + \sum_{\nu_2 \nu_3} D^{(1,1)}_{(1\nu_2\nu_3) 1;\alpha J}(\Omega^{-1})  \sum_{\tau \nu' I'} F^{\lambda}_{\tau}  
\ac^L_{1 1}(\lambda;\tau) D^{(\tau,\tau)}_{\nu' I';\bar\nu 1}(\Omega) \hat{T}^{\lambda}_{\tau;\nu' I'} \nonumber \\
&\quad +D^{(1,1)}_{(111) 0;\alpha J}(\Omega^{-1}) \sum_{\tau \nu' I'} F^{\lambda}_{\tau}  
\ac^L_{1 0}(\lambda; \tau) D^{(\tau,\tau)}_{\nu' I';\bar\nu 0}(\Omega) \hat{T}^{\lambda}_{\tau;\nu' I'}  . \label{explicitLA}
\eeqa

The right-hand side of Eq.(\ref{explicitLA}) contains $SU(3)$ $D$-functions  
{which are not  harmonic functions of the type of Eq.(\ref{HF}),
but which can be related to such functions by applying a differential operator 
transforming a coset function into an \emph{adjacent} function, differing from the coset function
by some $0, \pm 1$ changes in the occupation numbers.}  Details on obtaining these adjacency relations can be found in \ref{adjacencyrelations}.

The adjacency relations for $\nu_1=0,2$ and $I=\frac{1}{2}$ 
are similar to those for the $SU(2)$ $D$-functions \cite{Varshalovich} and 
can be compactly represented as 
\beq
\wz_{\nu \frac{1}{2}} 
D^{(\tau,\tau)}_{\mu J;(\tau\tau\tau)0}(\Omega)=
(-1)^{\nu_1/2}\sqrt{\frac{\tau(\tau+2)}{2}}D^{(\tau,\tau)}_{\mu J;\bar\nu \frac{1}{2}}(\Omega) \,  \label{12shift}, 
 \eeq
where {each} $\wz_{\nu \frac{1}{2}}$ is a first order differential operator
\beqa
\wz_{\nu \frac{1}{2}} = 
\sum_{k}d_{\nu \frac{1}{2}}(\Omega_k)
\frac{\partial }{\partial \Omega_k} \,  , 
\quad\; (\Omega_1,\Omega_2,\Omega_3,\Omega_4)=(\alpha_1,\beta_1,\alpha_2,\beta_2)\, .\label{S1}
\eeqa
The coefficients $d_{\nu \frac{1}{2}}(\Omega_k)$ are given in table \ref{tableofdcoefficients}.

To continue, we note that the expansion of Eq.(\ref{explicitLA}) contains terms 
of the type $D^{(\tau,\tau)}_{\nu'I';\bar\nu 1}(\Omega)$.  We argue here that these cannot be expressed as 
a first order differential operator acting on a coset function.  
To see how this fundamental difference with $SU(2)$ comes about, one must recognize  
that the action of any first order differential operator $\hat {\cal D}$
on a coset function \hdg{will drop operators linear in $\hat C_{ij}$ from the 
exponential $\hat R(\Omega)$; one can reorganize the resulting expression to}
\beqa  
\hat {\cal D}D^{(\tau,\tau)}_{\nu'I';(\tau\tau\tau)0}(\Omega)&=&
\sum_{i\ne j} c_{ij}
\bra{(\tau,\tau)\nu' I'}\hat R(\Omega) \hat C_{ij}\ket{(\tau,\tau)(\tau\tau\tau)0}\nonumber \\
&+& \bra{(\tau,\tau)\nu' I'}\hat R(\Omega) (c_1 \hat H_1+c_2\hat H_2)
\ket{(\tau,\tau)(\tau\tau\tau)0}\, .
\eeqa 
{by bringing any $\hat C_{ij}$  to the right of the $SU(3)$ rotation $\hat R(\Omega)$.  The action
of the $\mathfrak{su}(2)$ ladder operators $\hat C_{23}, \hat C_{32}$ kills $\ket{(\tau,\tau)(\tau\tau\tau)0}$, since by construction 
this state is an $\mathfrak{su}(2)$ scalar; the action of the remaining ladder operator $\hat C_{13},\hat C_{31}, \hat C_{12}$ and $\hat C_{21}$ 
will  lead to functions of the form $D^{(\tau,\tau)}_{\nu'I';\mu'\frac{1}{2}}(\Omega)$ by inspection, and thus not of the desired form.  Finally
the operators $\hat H_1$ and $\hat H_2$ both annihilate $\ket{(\tau,\tau)(\tau\tau\tau)0}$ by definition.}

{Instead, we find the adjacency relations for functions of the $D^{(\tau,\tau)}_{\nu'I';\bar\nu 1}(\Omega)$ type to be} of the form
\beqa
&&\wh_{1\nu_2\nu_3;1}D^{(\tau,\tau)}_{\mu J;(\tau\tau\tau)0}(\Omega) \nonumber \\
&&\qquad \qquad =-
{\tau(\tau+2)}\sqrt{\frac{(1+\delta_{\nu_2 1}\delta_{\nu_3 1})}{6}}
D^{(\tau,\tau)}_{\mu J;\bar \nu 1}(\Omega)\, ,
\label{S2ops}
\eeqa 
{where $\wh_{1\nu_2\nu_3;1}$ are} \emph{second order differential operators}
\beqa
 \wh_{(120);1}  &=
 \wz_{(021)\frac{1}{2}}\wz_{(210)\frac{1}{2}}
+ f_{(021)\frac{1}{2};(102)\frac{1}{2}}\wz_{(201)\frac{1}{2}} \nonumber \\
& +\textstyle\frac{\sqrt{2}}{2}f_{(021)\frac{1}{2};(111)1}\wz_{(210)\frac{1}{2}}
-\textstyle\sqrt{\frac{3}{2}}f_{(021)\frac{1}{2};(111)0}\wz_{(210)\frac{1}{2}} ,
 \label{S21201} \\
 \wh_{(102);1}&=\wz_{(012)\frac{1}{2}}\wz_{(201)\frac{1}{2}}
 - f_{(012)\frac{1}{2};(120)1}\wz_{(210)\frac{1}{2}} \nonumber \\
 & -
\left(\textstyle\frac{\sqrt{2}}{2} f_{(012)\frac{1}{2};(111)1}
+\textstyle\sqrt{\frac{3}{2}} f_{(012)\frac{1}{2};(111)0}\right)
\wz_{(201)\frac{1}{2}},
\label{S21021} \\
 -\frac{1}{\sqrt{3}} \wh_{(111);1} &
 =\wz_{021;\frac{1}{2}}\wz_{201;\frac{1}{2}} + \wz_{012;\frac{1}{2}}\wz_{210;\frac{1}{2}} \nonumber \\
& - f_{(021)\frac{1}{2};(120)1}\wz_{210;\frac{1}{2}} 
- f_{(012)\frac{1}{2};(102)1}\wz_{201;\frac{1}{2}} \nonumber \\
&- \left( \textstyle\sqrt{\frac{3}{2}}f_{(021)\frac{1}{2};(111)1} 
+ \textstyle\frac{\sqrt{2}}{2}f_{(021)\frac{1}{2};(111)0}\right)
 \wz_{201;\frac{1}{2}} \nonumber \\
 &\,
 - \left( \textstyle\sqrt{\frac{3}{2}}f_{(012)\frac{1}{2};(111)1} - \textstyle\frac{\sqrt{2}}{2}f_{(012)\frac{1}{2};(111)0}\right)
 \wz_{210;\frac{1}{2}} ,
 \label{eq:S21111}
\eeqa
{since they contain products of first order operators $\wz_{\mu\frac{1}{2}}$.}
The coefficients $f_{\beta a}$ are given in Table \ref{tableofsumsdc}. 

As a second step we use the relations Eqs.(\ref{12shift}) and (\ref{S2ops}) to explicitly determine the correspondence rules Eq.(\ref{explicitLA}), obtaining 
\beqa
&&\hat{T}^{\lambda}_{1;\alpha J} \wk:=\hat \mathfrak{C}_{\alpha J}^L \wk, \label{correspondenceshort}\\ 
&&\hat \mathfrak{C}_{\alpha J}^L= \sqrt{2}\sum_{\nu_1=0,2} \sum_{\nu_2\nu_3}
D^{(1,1)}_{\nu \frac{1}{2};\alpha J}(\Omega^{-1}) (-1)^{\bar \nu_1 /2} \wz_{\nu \frac{1}{2}}
\hat{\mathcal{C}}_2 ^{-1/2} \hat \ac^L_{ \nu_1 \frac{1}{2}}(\lambda;\hat{\mathcal{C}}_2) 
  \nonumber \\
&& \qquad  -  \sum_{\nu_2 \nu_3} D^{(1,1)}_{(1\nu_2\nu_3) 1;\alpha J}(\Omega^{-1})
\sqrt{\frac{6}{(1+\delta_{\nu_2 1}\delta_{\nu_3 1})}}
\wh_{1\nu_2\nu_3;1}
\hat{\mathcal{C}}_2 ^{-1} \hat \ac^L_{1 1}(\lambda;\hat{\mathcal{C}}_2)  \nonumber \\
&& \qquad + D^{(1,1)}_{(111) 0;\alpha J}(\Omega^{-1})   
\hat \ac^L_{ 1 0}(\lambda; \hat{\mathcal{C}}_2)  ,  \label{LA} 
\eeqa
where $\hat{\mathcal{C}}_2$ is  the differential realization of the $su(3)$ Casimir invariant,
\beqa
\hat{\mathcal{C}}_2 &=& \sum_{i\ne j} \hat{C}_{ij}\hat{C}_{ji} +\hat{H}_1^2+\hat{H}_2^2\, \\
\hat{\mathcal{C}}_2\,D^{(\sigma,\sigma)}_{\nu I;(\sigma\sigma\sigma)0}(\Omega)&=&\tau(\tau+2)D^{(\sigma,\sigma)}_{\nu I;(\sigma\sigma\sigma)0}(\Omega)\, ,
\label{C2}
\eeqa
on the harmonic functions Eq.(\ref{HF}).  This differential realization is given explicitly in \ref{sec:differentialCasimir}.  The operators $\hat \ac^L_{\nu_1 I}(\lambda;\hat {\cal C}_2)$ 
are functions  of the  $\hat{\mathcal{C}}_2$, such that
\begin{equation}
\hat \ac^L_{\mu_1 J}(\lambda;\hat{\mathcal{C}}_2)
D^{(\sigma,\sigma)}_{\nu I;(\sigma\sigma\sigma)0}(\Omega) 
=\ac^L_{\mu_1 J}(\lambda; \sigma )D^{(\sigma,\sigma)}_{\nu I;(\sigma\sigma\sigma)0}(\Omega) .
\end{equation}

The right action  
\beq 
\wk \hat{T}^{\lambda}_{1;\alpha J} := {\mathfrak C}^R_{\alpha J}\wk \label{eq:raction}
\eeq
is obtained by replacing 
$\ac^L_{\nu_1 I}(\lambda;\tau )   \rightarrow  \ac^R_{\nu_1 I}(\lambda; \tau)  $
and $\hat{\mathfrak C}^L_{\alpha J}\rightarrow  \hat {\mathfrak C}^R_{\alpha J}$ in Eq.(\ref{LA}), 
 with the coefficients $\ac^R_{\nu_1 I}(\lambda; \tau) $ 
and $\ac^L_{\nu_1 I}(\lambda;\tau)$ related quite simply by
\beqa
\ac^R_{0\frac{1}{2}}(\lambda;\tau)&= \ac^L_{0\frac{1}{2}}(\lambda;\tau) - \sqrt{3\tau(\tau+2)}
\, \nonumber\\
\ac^R_{11}(\lambda;\tau)&=\ac^L_{11}(\lambda;\tau)\, \nonumber\\
\ac^R_{10}(\lambda;\tau)&=\ac^L_{10}(\lambda;\tau)\, \nonumber \\
\ac^R_{2\frac{1}{2}}(\lambda;\tau)&=\ac^L_{2\frac{1}{2}}
(\lambda;\tau)+\sqrt{3\tau(\tau+2)}
\, .\label{br2half}
\eeqa

{The correspondence rules encapsulated in Eqs.(\ref{correspondenceshort}) and (\ref{eq:raction}) are the main results of 
this paper. The appearance of second order operators, 
$\wh_{1\nu_2\nu_3;1}$ , which are independent from the Casimir operator $\hat {\cal C}_2$,
is a key difference with the $SU(2)$ formalism, where the correspondence rules contain only first 
order differential operators and functions of the $SU(2)$ Casimir operator. 
This new "complication" for $SU(3)$ is ultimately a by-product of using
coset function $D^{(\tau,\tau)}_{\nu'I';(\tau\tau\tau)0}(\Omega)$ which are invariant under right action by an element in $U(2)$; because
the zero-weight subspace of $(\tau,\tau)$ will in general contain more than just an $I=0$ state, the action 
of generators with $I=1$ eventually yields $D^{(\tau,\tau)}_{\nu'I';\bar\nu1}(\Omega)$, which is no longer
$U(2)$-invariant under right action.  Such a situation 
cannot occur in $SU(2)$ since the $U(1)$-invariant zero-weight space is one-dimensional.
{Moreover}, it is clear that this complication will occur beyond $SU(3)$ for any $SU(n)$.}

Note that the coefficients $\ac^R$ and $\ac^L$ for $\nu_1=1$ do not change sign so that, as an immediate application 
of Eqs.(\ref{LA}) and (\ref{eq:raction}) we obtain using the relations 
of Eq. (\ref{br2half}) the commutator 
\beqa
[T^{\lambda}_{1;\alpha J},\wk ]
&= \frac{\sqrt{24}}{N}
\sum_{\nu}
D^{(1,1)}_{\nu\frac{1}{2};\alpha J}(\Omega^{-1})  \wz_{\nu\frac{1}{2}}
\wk \, ,
\label{Commw}
\eeqa
which 
contains only first order differential operators, in a manner similar to the $SU(2)$ case \cite{TWA}\cite{stratonovich}.

\section{Examples}

The  explicit form of the correspondence rules Eq.(\ref{LA}) can be used {to map} Schr\"{o}dinger equations for {Hamiltonians} polynomial in the generators  $\hat C_{ij}$ to evolution equations for the Wigner function.  Define
\beq 
\hat h=\sum_{\alpha, J} h_{\alpha J} \hat T^{\lambda}_{1,\alpha J}
\eeq 
and consider first Hamiltonians of the type 
$ \hat H=\hat h $ linear in the generators.  With $\hat \rho$ an arbitrary operators, 
we rewrite the symbol of the commutator as
\beqa
W_{[\hat H, \hat \rho]}(\Omega)=-\Tr(\hat \rho [\hat H,\wk]), 
\label{H1}
\eeqa
and use Eq.(\ref{symT}) together with the relation
\beq 
D^{(1,1)}_{\nu \frac{1}{2};\alpha J}(\Omega^{-1})=(-1)^{\nu_1/2} \, \sqrt{\frac{(\lambda+1)(\lambda+2)}{24}}\, 
\wz_{\nu;\frac{1}{2}}^* W_{\hat T^{\lambda}_{1;\alpha J}}(\Omega) 
\eeq 
to obtain from Eq.(\ref{Commw}) the following equation of motion for the Hamiltonian $\hat H=\hat h$: 
\beqa
 \partial _t W_{\hat \rho}(\Omega)=- \epsilon^{-1} \{ W_{\hat H}(\Omega), W_{\hat\rho} (\Omega)\}_ {\mathpzc{P}} 
\label{EqH1}
\eeqa
where 
\beqa
\epsilon^{-1}={2\sqrt{\lambda(\lambda+3)}}
\label{SP}
\eeqa
is the so-called semi-classical parameter,  and $\{\cdot , \cdot \}_{\mathpzc{P}} $
is the Poisson bracket given~\cite{hdg2011} by
\beq
\{ W_{\hat H}(\Omega), W_{\hat \rho}(\Omega) \}_{\mathpzc{P}} \nonumber \\
= -i \sum_{\nu}(-1)^{\frac{1}{2}\nu_1}
\left(
\wz_{\nu;\frac{1}{2}}^* W_{\hat H}(\Omega)\right)
\left(\wz_{\nu;\frac{1}{2}}
 W_{\hat\rho}(\Omega) \right) \, .
\label{PB}
\eeq
The sum in Eq.(\ref{PB}) is limited to $\nu_1=0,2$ as 
these are the only ones for which $I=\frac{1}{2}$.


Suppose next that $\hat H=\hat h^2$: the Hamiltonian is the square of a linear combination of $\mathfrak{su}(3)$ generators.   
Eq.(\ref{H1}) can be reduced to
\beqa
W_{[\hat h^2, \hat \rho]}=-\Tr(\hat \rho ([\hat h,\wk]\hat h+\hat h[\hat  h,\wk]))\, .
\eeqa 
In the specific case where $\hat H=(\hat T^{\lambda}_{1;\alpha J})^2$, we obtain from Eqs. (\ref{LA})  and (\ref{Commw}) 
\beqa
i \partial _t W_{\hat \rho}
= - \frac{\sqrt{24}}{N} 
\sum_{\nu}
D^{(1,1)}_{\nu\frac{1}{2};\alpha J}(\Omega^{-1})  \wz_{\nu\frac{1}{2}}
 (\hat \mathfrak{C}_{\alpha J}^L+\hat \mathfrak{C}_{\alpha J}^R) W_{\hat \rho} \, 
 \, .
\label{H2ex}
\eeqa
{Here again, the sum over $\nu$ is restricted to $\nu_1=0,2$ as only for those values can we have $I=\frac{1}{2}$.}

We emphasize that {Eq.(\ref{H2ex})}
contains third order derivatives (in addition to functions of the Casimir operator).  
{This is drastically different from the $SU(2)$ case, where third order differential operators appear only due to 
double derivatives in the differential expressions for 
Casimir functions. This third-order structure in the $SU(3)$ problem will lead to a very different long-time behaviour 
for Hamiltonians quadratic in $\mathfrak{su}(3)$ generators.}

In particular, for Hamiltonians $\hat H\propto (\hat T^{\lambda}_{1;(\alpha)111) 0})^2 $  and 
$\hat H\propto (\hat T^{\lambda}_{1;(\alpha)111) 1})^2$,  {the evolution} describes "$\mathfrak{su}(3)$" squeezing effect~\cite{su3SS}.
{The full expressions of Eq.(\ref{H2ex}) for these two cases can be found in \ref{sec:differentialoperators}.}
The evolution equation {for other cases is} quite complicated
{even for Hamiltonians $\hat H=\hat h^2$ quadratic in arbitrary combinations of generators.}

\section{Semiclassical  limit}

The semiclassical limit in quantum systems is associated with a large value of some physical 
parameter, {\it v.g.} the number of photons (in systems with $HW(1)$ symmetry), the size of an 
effective spin ($SU(2)$ symmetry), large value of a projection of angular momentum ($E(2)$ symmetry).  Then, 
the semiclassical expansion is performed in the inverse of this ``large'' parameter.
In physical realizations of quantum systems with $SU(3)$ symmetry, {\it v.g.} the Bose-Einstein 
condensate in a three-well configuration \hdg{\cite{BEstuff},} the semiclassical limit  would correspond to the 
large total number of excitations. From the mathematical perspective this corresponds to the 
large dimension of the (symmetric) representation. Then, $\epsilon$ 
defined in Eq.(\ref{SP}) (or  its approximate value $\epsilon \approx (2\lambda+3)^{-1}$) can be considered 
as appropriate semiclassical expansion parameter whenever $\epsilon \ll 1$.  

In order to obtain the asymptotic form of the 
correspondence rules we expand the operators  $\hat \ac^{R,L}_{ \nu I} (\lambda;\bar\nu)$ 
appearing in Eq.(\ref{LA}) in powers of $\epsilon$, keeping two non-vanishing orders,  {as given
explicitly in \ref{acoefficients}.} 
Then, the operators 
$\hat \mathfrak{C}^{L,R}$ take the following forms:
\beqa
\hat \mathfrak{C}_{\alpha J} ^{L,R} &=&N^{-1} \left[
\sqrt{6} \sum_{\nu_1=0,2}   \left(\pm 1+\textstyle\frac{3}{2}(-1)^{\nu_1/2} \epsilon \right)
\left( D^{(1,1)}_{\alpha J;\nu \frac{1}{2}}(\Omega)\right)^*
\wz_{\nu;\frac{1}{2}}\,
\right. \nonumber \\
&&\qquad - \epsilon\sum_{\nu_2\nu_3}
\sqrt{\frac{6}{(1+\delta_{\nu_21}\delta_{\nu_31})}} 
\left(D^{(1,1)}_{\alpha J;(1\nu_2\nu_3)1}(\Omega)\right)^*
\wh_{\nu;1}  \nonumber \\
  && \qquad\left. 
  +\left (  \frac{2}{\epsilon}-3 \epsilon (\hat{\mathcal{C}}_2 +3)\right )   \left(D^{(1,1)}_{\alpha J;(111)0}(\Omega)\right)^*
\right]\, .
\label{CLR}
\eeqa
with the $+$ and $-$ signs in the first term on the right for $\hat \mathfrak{C}_{\alpha J}^{L}$ and 
$\hat \mathfrak{C}_{\alpha J} ^{R}$, respectively. 
Then, we obtain from Eq. (\ref{H2ex})  the following approximate  equation of motion for a square of any 
$\mathfrak{su}(3)$ generator
\beqa
\partial _t W_{\hat \rho}= - \epsilon^{-1} \{ W_{\hat h^2}, W_{\hat \rho} \}_ {\mathpzc{P}} +O (\epsilon),
\label{eq:TWA} 
\eeqa
where no correction to the classical evolution of order $\epsilon^0$ in the semiclassical parameter 
 appears, as it is
expected for the Wigner function semiclassical dynamics. It is clear that
Eq.(\ref{eq:TWA}) corresponds to the so-called Truncated Wigner approximation \cite{TWA}, widely used in quantum systems with 
low-rank  
symmetries for the description of the semiclassical dynamic effects.

One should stress that there are two types of the second order differential operators, appearing in the first-order correction terms 
($\sim \epsilon$): the first type is proportional to the Casimir operator (this is similar to the $SU(2)$ situation) and the second type is
proportional to $\wh_{\nu;1}$: the  {appearance of this} latter type 
{qualitatively distinguishes the evolutions of systems with}
{low or higher rank symmetries.}

\section{Conclusions}

The correspondence rules of Eq.(\ref{LA}) can be immediately rewritten for the Wigner symbols and recast in terms of the star-product operation:
\beqa 
W_{\hat{T}^{\lambda}_{1;\alpha J} \hat \rho}&=\hat \mathfrak{C}_{\alpha J}^R 
W_{\hat \rho}=W_{\hat {T}^{\lambda}_{1;\alpha J} }
\star W_{\hat \rho}\, ,\\
W_{\hat \rho \hat  {T}^{\lambda}_{1;\alpha J} }&=\hat \mathfrak{C}_{\alpha,J}^L W_{\hat \rho}=W_{\hat \rho}
\star W_{\hat {T}^{\lambda}_{1;\alpha J} }.
\eeqa
The principal difference with other known correspondence rules consists in the appearance of second
order derivatives in the operators $\hat \mathfrak{C}_{\alpha J}^{L,R }$; this is in addition to the derivatives contained in
the differential form of the Casimir operators. 

Apart from derivatives arising in the Casimir operator, the exact  equation of motion describing the non-linear evolution of the Wigner 
function contains at least third-order differential operators; this significantly  complicates the analysis of non-linear dynamics in 
comparison, for instance,  with spin evolution.  

\andrei{In spite of this novel feature, 
the leading order term of the semiclassical expansion
of the evolution equation is still reduced to the Poisson brackets on the $\mathbb{CP}^2$-manifold,
and the short-time dynamics can still be well described in terms of classical trajectories.  
The appearance of third order derivatives in the exact equations of motion
significantly affects the qualitative character of non-linear evolution beyond the semiclassical times.   The explicit forms
of terms of order $\epsilon$ and $1/\epsilon$ in Eq.(\ref{CLR}) - terms that cancel to leading order in the 
semi-classical expansion of the quantum Hamiltonian evolution - are 
non-trivial and quite important {\it e.g.} for the phase-space description of $SU(3)$ dissipative channels.}

As an interesting and important by-product we have obtained adjacency relations connecting some $SU(3)$ $D$-functions to the 
harmonic functions on $SU(3)/U(2)$, thereby generalizing results on spherical harmonics 
{(see Chapter 4 of \cite{Varshalovich}). }
We expect to use these novel relations in application of the $SU(3)$ group in quantum mechanics.

We have also formally shown that a single parameter $\epsilon$, given in Eq.(\ref{SP}), 
related to the inverse eigenvalue of the Casimir, and which scales like the inverse of the label $\lambda$ of the irrep
$(\lambda,0)$,  naturally enters in the semiclassical limit of the nonlinear $SU(3)$ dynamics

HdG acknowledges the support of NSERC of Canada for this work;  the work of ABK is supported by CONACyT grant 254127.

\appendix 

\section{Technical results on Clebsch, tensors and Racah}


The $SU(3)$ Clebsch-Gordan coefficients factor into a reduced (or double-barred) 
coefficient multiplied by a $SU(2)$ coefficient:
\beqa 
\kern-2em \cg{(\lambda_1,\mu_1)}{(a_1a_2a_3) I_1}{(\lambda_2,\mu_2)}{(b_1b_2b_3) I_2}{(\lambda,\mu)}{(c_1c_2c_3) J}_\kappa
&=&\rcg{(\lambda_1,\mu_1)}{a_1 I_1}{(\lambda_2,\mu_2)}{b_1 I_2}{(\lambda,\mu)}{\mu_1 J}_\kappa \nonumber \\
&\times& \cg{I_1}{\frac{1}{2}(a_2-a_3)}{I_2}{\frac{1}{2}(b_2-b_3)}{J}{\frac{1}{2}(c_2-c_3)}
\eeqa 
where $\kappa$ labels (where appropriate) the multiple copies of the irrep $(\lambda,\mu)$ in the decomposition of the tensor product 
$(\lambda_1,\mu_1)\otimes (\lambda_2,\mu_2)$.  This index $\kappa$ is omitted when $(\lambda,\mu)$ occurs once in 
$(\lambda_1,\mu_1)\otimes (\lambda_2,\mu_2)$.  

\subsection{Tensor operators}\label{sec:tensoroperators}

One can show \cite{CGAlex} that
\beqa 
&\rcg{(\lambda,0)}{\lambda-a;\frac{1}{2}a}{(0,\lambda)}{\sigma+a;\frac{1}{2}(\sigma+a)}{(\sigma,\sigma)}{2\sigma;\frac{1}{2}\sigma}
\nonumber \\
&\qquad =(-1)^a \sqrt{\frac{(\sigma+a+1)!(\lambda-\sigma)!(\lambda-a)!(2\sigma+2)!}{(\sigma+1)!(\lambda-\sigma-a)!a!\sigma!(\lambda+\sigma+2)!}}
\eeqa 
for the highest weight state, and more generally, for any state in $(\sigma,\sigma)$
\beqa 
 &\rcg{(\lambda,0)}{\nu_1';\frac{1}{2}(\lambda-\nu_1')}{(0,\lambda)}{\lambda+\sigma-\nu_1'-p;\frac{1}{2}(\lambda+\sigma-\nu_1'-p)}{(\sigma,\sigma)}{2\sigma-p;I}=(-1)^\lambda\,\frac{(\lambda-\nu_1')!}{\sigma!} \nonumber\\
&\times
\textstyle\sqrt{\frac{(\sigma-I+\half(\sigma-p))!(\sigma+I+\half(\sigma-p)+1)!(\lambda-\sigma)!(2\sigma+2)(\nu_1'+p-\sigma)!(1+I+\half(p+\sigma))!}{(2+\lambda+\sigma)!(\nu_1')!(\lambda-\nu_1'-I+\half(\sigma-p))!(1+I+\lambda-\nu_1'+\half(\sigma-p))!(\half(\sigma+p)-I)!}}\nonumber\\
&\times\sum_{\nu_1=\nu_{1,\min}}^{\nu_{1,\max}}(-1)^{\nu_1}\Big(\frac{(\lambda-I-\nu_1+\half(\sigma+p))!\nu_1!}{(\nu_1-\nu_1')!(p-\nu_1+\nu_1')!(\nu_1-\sigma)!(\lambda-\nu_1)!}\Big)\nonumber\\
 &\times\textstyle{_3}F_2
 \left[
 {\renewcommand{\arraystretch}{1.3}
 \renewcommand{\arraycolsep}{4.85pt}\begin{array}{lll}
 \nu_1-\lambda& I+\nu_1'-\lambda+\frac{1}{2}(p-\sigma)&1+I+\frac{1}{2}(\sigma-p) \\
 \nu_1'-\lambda &I+\nu_1-\lambda-\frac{1}{2}(p+\sigma)
 \end{array}};1\right]\label{eq:74}
\eeqa
with $_3F_2$ the generalized hypergeometric function, and
\beq 
\nu_{1,\min}=\max[\sigma,\nu_1']\, ,\; \nu_{1,\max}=\min[\lambda,\nu_1'+p,\lambda-I+\half(\sigma+p)]
\eeq 
The bounds of the sum are determined by the factorials involved with $\nu_1$.

The combinations
\beqa 
\hat T^{\lambda}_{\sigma;\nu I_\nu}&=&
\sum_{\alpha I_\alpha \beta I_\beta}\tilde{C}^{\sigma \nu I_{\nu}}_{\lambda \alpha I_\alpha;\lambda^* \beta I_{\beta}}
\ket{(\lambda,0)\alpha I_\alpha}\bra{(\lambda,0)\beta I_\beta}\, ,\label{eq:tensordef}\\
\tilde{C}^{\sigma \nu I_{\nu}}_{\lambda \alpha I_\alpha;\lambda^* \beta^* I_{\beta}}
&=&\cg{(\lambda,0)}{\alpha I_\alpha }{(0,\lambda)}{\beta^* I_\beta}{(\sigma,\sigma)}{\nu I_\nu} 
(-1)^{\lambda-\beta_2}\label{eq:ctildetensor}
\eeqa 
with $\beta^*=(\lambda-\beta_1,\lambda-\beta_2,\lambda-\beta_3)$, 
are $\mathfrak{su}(3)$ irreducible tensor operators.


The irreducible tensor operators of equation (\ref{deftensorope}) satisfy the following trace-orthogonality condition 
\beqa
\Tr((\hat{T}^\lambda_{\sigma;\alpha I_\alpha})^\dagger\hat{T}^\lambda_{\sigma';\alpha' I_{\alpha'}})=\delta_{\sigma \sigma'}\delta_{\alpha \alpha'}\delta_{I_\alpha I_{\alpha'}},
\label{orthogonalityrelation}
\eeqa
where
\beqa
\hat{T}^\lambda_{\sigma;\nu J}=(-1)^{\sigma+\nu_2}(\hat{T}^{\lambda}_{\sigma;\nu^* J})^\dagger
\label{daggertensor}
\eeqa
with $\nu^*=(2\sigma-\nu_1,2\sigma-\nu_2,2\sigma-\nu_3)$. 

One can decompose products like $T^\lambda_{1;\nu I}T^{\lambda}_{\sigma;(\sigma\sigma\sigma)0}$ using 
the definition of the Racah $U$-coefficients \cite{DraayerAkiyama}
and the orthogonality property of tensors:
\beqa
&&\hat T^{(1,1)}_{\nu;J} T^{(\sigma,\sigma)}_{(\sigma\sigma\sigma)0}=\sum_{\tau=\sigma-1}^{\sigma+1}
c_{\nu_1J;\sigma}^{(\tau,\tau)}\hat T^{(\tau,\tau)}_{\nu';J}\, ,\\
&&c_{\nu_1J;\sigma}^{(\tau,\tau)}=\sum_\rho 
\rcg{(1,1)}{\nu_1;J}{(\sigma,\sigma)}{\sigma;0}{(\tau,\tau)}{\nu_1';J}_\rho
U_\rho[(1,1)(\lambda,0)(\tau,\tau)(0,\lambda);(\lambda,0)(\sigma,\sigma)] \nonumber 
\eeqa 
Here we provide tables of reduced CG coefficients, Racah $U$-coefficients and others coefficients needed to obtains some intermediate results
or otherwise useful in calculations.

\subsection{Some CG for couplings of the type $(1,1)\otimes (\sigma,\sigma)
\to (\tau,\tau)$ }

The highest weight for the irrep $(\sigma+1,\sigma+1)$ in the reduction of the product $(1,1)\otimes (\sigma,\sigma)$ 
is the product of the $(1,1)$ and $(\sigma,\sigma)$ highest weights so that 
\beqa
\rcg{(1,1)}{2;\frac{1}{2}}{(\sigma,\sigma)}{(2\sigma;\frac{1}{2}\sigma)}{(\sigma+1,\sigma+1)}{2(\sigma+1);\frac{1}{2}(\sigma+1)}=1
\, .
\label{11sss+1s+1}
\eeqa

\begin{table}[H]
  \begin{center}
  \renewcommand*{\arraystretch}{2.25}
  \caption{$\rcg{(1,1)}{\nu_1;I}{(\sigma,\sigma)}{\sigma;0}{(\sigma+1,\sigma+1)}{\mu;I}$}
  \label{tableofhighestweights+1}
    \begin{tabular}{|c|c|c|} 
    \hline
    $(\nu_1;I)$&$(\mu;I)$ & $\rcg{(1,1)}{\nu_1;I}{(\sigma,\sigma)}{\sigma;0}{(\sigma+1,\sigma+1)}{\mu;I}$ \\
    \hline
    $(2;\frac{1}{2})$&$(\sigma+2;\frac{1}{2})$ & $\displaystyle\frac{(\sigma+2)}{2(\sigma+1)}\sqrt{\frac{(\sigma+3)}{(2\sigma+3)}}
$ \\
\hline
$(1;1)$&$(\sigma+1;1)$& $\displaystyle\frac{(\sigma+2)(\sigma+3)}{2(\sigma+1)}\sqrt{\frac{1}{3(\sigma+1)(2\sigma+3)}}$ \\
\hline
$(1;0)$&$(\sigma+1;0)$& $\displaystyle\frac{(\sigma+2)}{2}
\sqrt{\frac{3}{(2\sigma+3)(\sigma+1)}}$ \\
\hline
$(0;\frac{1}{2})$&{$(\sigma;\frac{1}{2})$}&$\displaystyle\frac{(\sigma+2)}{2(\sigma+1)}\sqrt{\frac{\sigma+3}{(2\sigma+3)}}$ \\
\hline
    \end{tabular}
  \end{center}
\end{table}

\begin{table}[H]
  \begin{center}
  \renewcommand*{\arraystretch}{2.25}
  \caption{$\rcg{(1,1)}{\nu_1;J}{(\sigma,\sigma)}{n_1;I_n}{(\sigma-1,\sigma-1)}{2(\sigma-1);\frac{1}{2}(\sigma-1)}$}
    \begin{tabular}{|c|c|c|} 
    \hline
    $(\nu_1;J)$&$(n_1;I_n)$& $\rcg{(1,1)}{\nu_1;J}{(\sigma,\sigma)}{n_1;I_n}{(\sigma-1,\sigma-1)}{2(\sigma-1);\frac{1}{2}(\sigma-1)}$\\
    \hline
    $(2;\frac{1}{2})$&$(2(\sigma-1);\frac{1}{2}\sigma-1)$& $-\frac{1}{\sigma+1}\sqrt{\frac{(\sigma-1)}{(2\sigma+1)}}$\\
    \hline
    $(2;\frac{1}{2})$&$(2(\sigma-1);\frac{1}{2}\sigma)$& $\frac{1}{\sigma+1}\sqrt{\frac{\sigma+2}{2\sigma+1}}$ \\
    \hline
    $(1;1)$&$(2\sigma-1;\frac{1}{2}(\sigma-1))$ & $\frac{\sigma}{\sigma+1}\sqrt{\frac{(\sigma-1)}{2(\sigma+1)(2\sigma+1)}}$ \\
    \hline
    $(1;1)$&$(2\sigma-1;\frac{1}{2}(\sigma+1))$ & $-\frac{1}{\sigma+1}\sqrt{\frac{\sigma(\sigma+2)}{(\sigma+1)}}$ \\
    \hline
    $(1;0)$&$(2\sigma-1;\frac{1}{2}(\sigma-1))$ & $\frac{\sigma}{\sigma+1}\sqrt{\frac{3}{2(2\sigma+1)}}$ \\
  \hline
    \end{tabular}
  \end{center}
\end{table}

\begin{table}[H]
  \begin{center}
  \renewcommand*{\arraystretch}{2.5}
  \caption{$\rcg{(1,1)}{\nu_1;J}{(\sigma,\sigma)}{\sigma;0}{(\sigma-1,\sigma-1)}{\mu;J}$}
    \begin{tabular}{|c|c|c|} 
      \hline
    $(\nu_1;J)$&$(\mu;J)$& $\rcg{(1,1)}{\nu_1;J}{(\sigma,\sigma)}{\sigma;0}{(\sigma-1,\sigma-1)}{\mu;J}$\\
    \hline
    $(2;\frac{1}{2})$&$(\sigma;\frac{1}{2})$& $-\frac{\sigma}{2(\sigma+1)}\sqrt{\frac{(\sigma-1)}{(2\sigma+1)}}$\\
    \hline
    $(1;1)$&$(\sigma-1,1)$& $\frac{\sigma(\sigma-1)}{2(\sigma+1)\sqrt{3(\sigma+1)(2\sigma+1)}}$\\
    \hline
       $(1;0)$&$(\sigma-1;0)$ & $\frac{\sigma}{\sigma+1}\sqrt{\frac{3}{2(2\sigma+1)}}$ \\
  \hline
    $(0;\frac{1}{2})$&$(\sigma-2;\frac{1}{2})$& $\frac{\sigma}{2}\,
\sqrt{\frac{3}{(2\sigma+1)(\sigma+1)}} $  \\
    \hline
    \end{tabular}
  \end{center}
\end{table}
Some calculations also require the following expressions: 
\beqa
&&\rcg{(1,1)}{1;0}{(\sigma,\sigma)}{2\sigma-1-p;I}{(\sigma-1,\sigma-1)}{2(\sigma-1)-p;I}\nonumber \\
&&\qquad =\frac{(-1)^{p-2I+\sigma+1}}{8(\sigma+1)}\sqrt{\frac{3(3+2I+p+\sigma)(1-2I+p+\sigma)}{(\sigma+1)(2\sigma+1)}} \nonumber \\
&&\qquad \times\sqrt{(1+2I-p+3\sigma)(-1-2I-p+3\sigma)} \\
&&\rcg{(1,1)}{1;0}{(\sigma,\sigma)}{2(\sigma+1)-p-1;I}{(\sigma+1,\sigma+1)}{2(\sigma+1)-p;I}\nonumber \\
&&\qquad =\frac{(-1)^{p-2I+\sigma+1}}{2(\sigma+1)}\sqrt{\frac{3(3+2I+p+\sigma)(1-2I+p+\sigma)}{(\sigma+1)(2\sigma+3)}}\nonumber \\
&&\qquad \times\sqrt{(3-2I-p+3\sigma)(5+2I-p+3\sigma)}
\eeqa

\begin{table}[H] \centering
\caption{The $SU(3)$ reduced CG 
$\rcg{(1,1)}{\nu_1 I_1}{(\sigma,\sigma)}{n_1I_2}
{(\sigma,\sigma)}{2\sigma;\frac{1}{2}\sigma}_\rho$. 
The $\rho=1$ copy is chosen using the usual convention that the $SU(3)$ CGs 
agree with the Wigner-Eckart theorem when the generators are considered as $SU(3)$ tensors transforming by the $(1,1)$ representation. The  $\rho=2$
copy is chosen to be orthogonal to the $\rho=1$ copy.}
{\renewcommand{\arraystretch}{1.75}
\begin{tabular}{|C|C|C|C|} 
\hline 
\nu_1;I_1&n_1;I_2\phantom{\begin{array}{c}{\scriptsize{a}}\\ {\scriptsize{a}}\end{array}}& \rcg{(1,1)}{\nu_1 I_1}{(\sigma,\sigma)}{n_1I_2}
{(\sigma,\sigma)}{2\sigma;\frac{1}{2}\sigma}_1 & \rcg{(1,1)}{\nu_1 I_1}{(\sigma,\sigma)}{n_1I_2}
{(\sigma,\sigma)}{2\sigma;\frac{1}{2}\sigma}_2\\
\hline 
1;1&2\sigma;\frac{1}{2}\sigma
&  \frac{1}{2} & -\frac{\sqrt{3}}{2}\sqrt{\frac{2\sigma+1}{2\sigma+3}} \\
\hline
1;0&2\sigma;\frac{1}{2}\sigma
& \frac{1}{2}\sqrt{\frac{3\sigma}{\sigma+2}} & \frac{1}{2}\sqrt{\frac{\sigma(2\sigma+1)}{(\sigma+2)(2\sigma+3)}} \\
\hline 
2;\frac{1}{2}&2\sigma-1,\frac{1}{2}(\sigma+1)&
\sqrt{\frac{\sigma+2}{2(\sigma+1)(\sigma+2)}} & \sqrt{\frac{3(2\sigma+1)}{2(\sigma+1)(2\sigma+3)}} \\
\hline 
2;\frac{1}{2}&2\sigma-1;\frac{1}{2}(\sigma-1)&
-\sqrt{\frac{(2\sigma+1)}{2(\sigma+1)(\sigma+2)}} & 
\sqrt{\frac{3}{2(\sigma+1)(\sigma+2)(2\sigma+3)}} \\ 
\hline
\end{tabular}
}
\label{copiesrho1rho2}
\end{table}

\begin{table}[H] \centering
\caption{The $SU(3)$ reduced CG 
$\rcg{(1,1)}{\nu_1 I_1}{(\sigma,\sigma}{\sigma;0}
{(\sigma,\sigma)}{N_1 J}_\rho$. }
{\renewcommand{\arraystretch}{1.75}
\begin{tabular}{|C|C|C|C|} 
\hline 
\nu_1;I_1&N_1;J\phantom{\begin{array}{c}a\\b\end{array}}& 
\rcg{(1,1)}{\nu_1 I_1}{(\sigma,\sigma}{\sigma;0}{(\sigma,\sigma)}{N_1 J}_1 & 
\rcg{(1,1)}{\nu_1 I_1}{(\sigma,\sigma}{\sigma;0}{(\sigma,\sigma)}{N_1 J}_2\\
\hline 
2;\frac{1}{2}&\sigma+1;\frac{1}{2}
&  -\frac{1}{2} &\sqrt{\textstyle\frac{3}{4(2\sigma+1)(2\sigma+3)}} \\
\hline
1;0&\sigma;0
& 0 &\sqrt{\frac{\sigma(\sigma+2)}{(2\sigma+1)(2\sigma+3)}} \\
\hline 
1;1&\sigma;1&
0& -\sqrt{\textstyle\frac{\sigma(\sigma+2)}{(2\sigma+1)(2\sigma+3)}} \\
\hline 
0;\frac{1}{2}&\sigma-1;\frac{1}{2}&
\frac{1}{2} & 
\textstyle\frac{\sqrt{3}}{2}\textstyle\frac{1}{\sqrt{(2\sigma+1)(2\sigma+3)}}\\ 
\hline
\end{tabular}
}
\end{table}

Some final analytical expressions for the decomposition $(1,1)\otimes (\sigma,\sigma)\rightarrow (\sigma,\sigma)$ are also needed:
\beqa
&&\rcg{(1,1)}{1;0}{(\sigma,\sigma)}{2\sigma-p;I}{(\sigma,\sigma)}{2\sigma-p;I}_{\rho=1}\nonumber \\
&&=\frac{\sqrt{3}(-1)^{p+\sigma-2I}}{2(\sigma+1)\sqrt{\sigma(\sigma+2)}}\Big(\sigma(\sigma+1)
-\textstyle\frac{1}{4}(p-\sigma+2I) \nonumber \\
&&\qquad \times(\sigma-p+2I+2)-\textstyle\frac{1}{4}(p+\sigma-2I)(p+\sigma+2I+2)\Big) \\
&&\rcg{(1,1)}{1;0}{(\sigma,\sigma)}{2\sigma-p;I}{(\sigma,\sigma)}{2\sigma-p;I}_{\rho=2}\nonumber \\
&&=\frac{(-1)^{p+\sigma-2I}\sqrt{2\sigma+1}}{2(\sigma+1)\sqrt{\sigma(\sigma+2)(2\sigma+3)}}\Big(\sigma(\sigma+1)
\nonumber \\
&&\; -\textstyle\frac{3}{4}(p-\sigma+2I) (\sigma-p+2I+2)+{\textstyle\frac{3}{4}}
\frac{(p+\sigma-2I)(p+\sigma+2I+2)}{(2\sigma+1)}\Big)
\eeqa

\subsection{Basic equation to evaluate $U_{\hbox{\tiny su(3)}}$}

\begin{table}[H] \centering
\caption{The Racah U-coefficients}
\label{tableofRacahU}
{\renewcommand{\arraystretch}{1.6}
\begin{tabular}{| C | C | C |}
\hline
\tau &\rho & U_{\rho}\left[(1,1)(\lambda,0)(\tau,\tau)(0,\lambda);(\lambda,0)(\sigma,\sigma)\right]_\rho\\
\hline
\sigma+1 & &\frac{(\sigma+1)}{2}\sqrt{\frac{3(\lambda-\sigma)(\lambda+\sigma+3)}{\lambda(\lambda+3)(\sigma+2)(2\sigma+3)}}\\
\hline
\sigma & 1 & \frac{\sqrt{3}}{2}\sqrt{\frac{\sigma(\sigma+2)}{\lambda(\lambda+3)}}\\
\hline
\sigma & 2 & \frac{(2\lambda+3)}{2}\sqrt{\frac{\sigma(\sigma+2)}{\lambda(\lambda+3)(2\sigma+1)(2\sigma+3)}}\\
\hline
\sigma-1 & &-\frac{(\sigma+1)}{2}\sqrt{\frac{3(\lambda-\sigma+1)(\lambda+\sigma+2)}{\lambda(\lambda+3)\sigma(2\sigma+1)}} \\
\hline
\end{tabular}
}
\end{table}

The basic equation to evaluate $U_{\hbox{\tiny su(3)}}$ coefficients is \cite{DraayerAkiyama}:
\beqa
&&
\kern -1em \sum_{\rho}\rcg{(1,1)}{\nu_1; I}{(\sigma,\sigma)}{2\sigma ; \frac{1}{2}\sigma}{(\tau,\tau)}{2\tau; \frac{1}{2}\tau}_\rho\,
U_{\hbox{\tiny su(3)}}\left((1,1),(\lambda,0),(\tau,\tau),(0,\lambda);(\lambda,0),(\sigma,\sigma)\right)_\rho\nonumber \\
&&=\sum_{a_1}\rcg{(\lambda,0)}{a_1}{(0,\lambda)}{\lambda+\sigma-a_1}{(\sigma,\sigma)}{2\sigma;\frac{1}{2}\sigma}\,
\rcg{(1,1)}{\nu_1;I}{(\lambda,0)}{a_1}{(\lambda,0)}{\nu_1-1+a_1}\nonumber\\
&&\qquad\times \rcg{(\lambda,0)}{\nu_1-1+a_1}{(0,\lambda)}{\lambda+\sigma-a_1}{(\tau,\tau)}{2\tau;\frac{1}{2} \tau}\,
U_{\hbox{\tiny su(2)}}\left(I,I_a,\textstyle\frac{1}{2} \tau, I_b;I_c,\hf\sigma\right)
\label{basicU}
\eeqa
where $\rho$ labels the copies of the irrep $(\tau,\tau)$ in the decomposition of the product $(1,1)\otimes(\sigma,\sigma)$.  Specifically,
the irreps $(\sigma+1,\sigma+1)$ and $(\sigma-1,\sigma-1)$ occur once so $\rho$ is redundant, but the irrep $(\sigma,\sigma)$ occurs
twice so when $(\tau,\tau)=(\sigma,\sigma)$ there is a sum of the two copies of this irrep.

In addition, we have:
\beq
I_a=\hf(\lambda-a_1)\, ,\quad I_b=\hf(\lambda+\sigma-a_1)\, ,\quad I_c=\hf(\lambda-\nu_1-a_1+1)\,  .
\eeq

In Eq.(\ref{basicU}), the indices $b^*$ and $c$ are implicit and related to $a$ and $\nu$ through
\beq
\begin{array}{lll}
b_1^*=\lambda+\sigma-a_1\, ,\quad & b_2^*=\lambda-a_2\, ,\quad &b_3=\lambda-\sigma-a_3\\
c_1=\nu_1-1+a_1\, ,& c_2=\nu_2-1+a_2\, &c_3=\nu_3-1+a_3
\end{array}
\eeq

On the right  hand side, we need CGs of the type given in Table (\ref{tableofhighestweights+1}).
The Racah $U_{su(2)}$ coefficient  is related to the Wigner $6j$-symbol by
\beq
U(abcd;ef)=(-1)^{a+b+c+d}\,\sqrt{(2e+1)(2f+1)}\,\sixj{a}{b}{e}{d}{c}{f}\, .
\eeq

Reduced CG's of the type $\rcg{(1,1)}{\nu_1;I}{(\lambda,0)}{a_1}{(\lambda,0)}{\nu_1-1+a_1}$ can be obtained
from the algorithm of \cite{CGAlex} and depend on three cases, which are also tied 
to the relation between $\tau$ and $\sigma$.  It is also useful to note the symmetry relation
\beqa
&&\cg{(\lambda,0)}{\alpha_1\alpha_2\alpha_3}{(0,\lambda)}{\lambda-\beta_1,\lambda-\beta_2,\lambda-\beta_3}{(1,1)}{\gamma_1\gamma_2\gamma_3;I}
\nonumber \\
&&=(-1)^{\beta_2}\,\sqrt{\frac{16}{(\lambda+1)(\lambda+2)}}\,
\cg{(1,1)}{\gamma_1\gamma_2\gamma_3;I}{(\lambda,0)}{\beta_1\beta_2,\beta_3}{(\lambda,0)}{\alpha_1\alpha_2\alpha_3}
\label{eq:symmetryrelation}
\eeqa
which can be obtained from \cite{CGAlex}.

\subsubsection{$\nu_1=2$}

If $\nu_1=2$, then $I'=\hf$ and $\tau=\sigma+1$.  We can combine Eq.(\ref{11sss+1s+1}), use Eq.(\ref{eq:symmetryrelation})
and the tables already provided to obtain
\beqa
&&\rcg{(1,1)}{2;\frac{1}{2}}{(\lambda,0)}{a_1}{(\lambda,0)}{a_1+1}
=\sqrt{\frac{3(\lambda-a_1+1)(a_1+1)}{2\lambda(\lambda+3)}}\, ,\\ 
&&U_{\hbox{\tiny su(2)}}\left(\hf,I_a,\hf \sigma+\hf, \hf(\lambda+\sigma-a_1);I_a-\hf,\hf\sigma\right)\nonumber \\
&&\qquad \qquad= - \sqrt{\frac{(\lambda-a_1)(\sigma+1)}{(\lambda-a_1+1)(\sigma+2)}}\,  .
\eeqa
and evaluate the sum on the right of Eq.(\ref{basicU}) as
\beqa
&&U_{\hbox{\tiny su(3)}}\left((1,1),(\lambda,0),(\sigma+1,\sigma+1),(0,\lambda);(\lambda,0),(\sigma,\sigma)\right)\nonumber \\ 
&&\qquad\qquad =
\frac{ (\sigma +1)}{2}  \sqrt{ \frac{3(\lambda -\sigma ) (\lambda +\sigma +3)}{\lambda  (\lambda +3) (\sigma +2) (2 \sigma +3)}}
\eeqa

\subsubsection{$\nu_1=1$}

When $\nu_1=1$, we have $I'=0$ or $I'=1$, and $\tau=\sigma$. For $I'=0$ we have:
\beqa 
\rcg{(1,1)}{1;0}{(\lambda,0)}{a_1}{(\lambda,0)}{a_1}
&=&
\frac{3a_1-\lambda}{2\sqrt{\lambda(\lambda+3)}}\nonumber \\
U_{\hbox{\tiny su(2)}}\left(0,I_a,\textstyle\frac{1}{2}\sigma, \textstyle\frac{1}{2}(\lambda+\sigma-a_1);I_a,\hf\sigma\right)&=&1\, ,\nonumber\\
\rcg{(1,1)}{1; 0}{(\sigma,\sigma)}{2\sigma ; \frac{1}{2}\sigma}{(\sigma,\sigma)}{2\sigma; \frac{1}{2}\sigma}_{\rho=1}
&=&{\displaystyle\frac{\sqrt{3}}{2}}\sqrt{\frac{\sigma}{(\sigma+2)}}\, , \nonumber\\
\rcg{(1,1)}{1; 0}{(\sigma,\sigma)}{2\sigma ; \frac{1}{2}\sigma}{(\sigma,\sigma)}{2\sigma; \frac{1}{2}\sigma}_{\rho=2}
&=&{\displaystyle\frac{1}{2}}\sqrt{\frac{\sigma(2\sigma+1)}{(\sigma+2)(2\sigma+3)}}\, .
\eeqa
The sum on RHS of Eq.(\ref{basicU}) can here again be evaluated in closed form to produce
\beqa
&&
\frac{\sqrt{3}}{2}
\sqrt{\frac{\sigma}{\sigma+2}}\,
U_{\hbox{\tiny su(3)}}\left((1,1),(\lambda,0),(\sigma,\sigma),(0,\lambda);(\lambda,0),(\sigma,\sigma)\right)_{\rho=1}
\nonumber\\
&&\kern -1em +\sqrt{\frac{\sigma(2\sigma+1)}{4(\sigma+2)(2\sigma+3)}}
U_{\hbox{\tiny su(3)}}\left((1,1),(\lambda,0),(\sigma,\sigma),(0,\lambda);(\lambda,0),(\sigma,\sigma)\right)_{\rho=2}\nonumber \\
&&=\frac{\sigma(\lambda+3\sigma+6)}{2(2\sigma+3)\sqrt{\lambda(\lambda+3)}} \, .
\label{UeqIequals0}
\eeqa

For $I'=1$, we find
\beqa
\rcg{(1,1)}{1;1}{(\lambda,0)}{a_1}{(\lambda,0)}{a_1}&=&
\frac{1}{2}\sqrt{\frac{3(\lambda-a_1)(\lambda-a_1+2)}{\lambda(\lambda+3)}}\, ,\nonumber \\
U_{\hbox{\tiny su(2)}}\left(1,I_a,\textstyle\frac{1}{2}\sigma, \textstyle\frac{1}{2}(\lambda+\sigma-a_1);I_a,\hf\sigma\right)
&=&-\sqrt{\frac{(\lambda-a_1)\sigma}{(\sigma+2)(\lambda-a_1+2)}}\, , \nonumber \\
\rcg{(1,1)}{1; 1}{(\sigma,\sigma)}{2\sigma ; \frac{1}{2}\sigma}{(\sigma,\sigma)}{2\sigma; \frac{1}{2}\sigma}
&=&{\textstyle\frac{1}{2}}\, . 
\eeqa
and this time
\beqa
&&\frac{1}{2}U_{\hbox{\tiny su(3)}}\left((1,1),(\lambda,0),(\sigma,\sigma),(0,\lambda);(\lambda,0),(\sigma,\sigma)\right)_{\rho=1}
\nonumber \\
&&-\frac{\sqrt{3}}{2}\sqrt{\frac{2\sigma+1}{2\sigma+3}}
U_{\hbox{\tiny su(3)}}\left((1,1),(\lambda,0),(\sigma,\sigma),(0,\lambda);(\lambda,0),(\sigma,\sigma)\right)_{\rho=2}\nonumber \\
&&\qquad =-\frac{(\lambda-\sigma)}{2(2\sigma+3)}\sqrt{\frac{3\sigma(\sigma+2)}{\lambda(\lambda+3)}}
\label{UeqIequals1}
\eeqa
This system is inverted to obtain the final expressions
\beqa
&&U_{\hbox{\tiny su(3)}}\left((1,1),(\lambda,0),(\sigma,\sigma),(0,\lambda);(\lambda,0),(\sigma,\sigma)\right)_{\rho=1}
 \nonumber \\
 &&\qquad \qquad \qquad 
=\frac{\sqrt{3}}{2}\sqrt{\frac{\sigma(\sigma+2)}{\lambda(\lambda+3)}}\, ,\\
&&U_{\hbox{\tiny su(3)}}\left((1,1),(\lambda,0),(\sigma,\sigma),(0,\lambda);(\lambda,0),(\sigma,\sigma)\right)_{\rho=2} \nonumber \\
&&\qquad\qquad \qquad  = \frac{(2\lambda+3)}{2}\sqrt{\frac{\sigma(\sigma+2)}{\lambda(\lambda+3)(2\sigma+1)(2\sigma+3)}}\, .
\eeqa

\subsubsection{$\nu_1=0$}

When $\nu_1=0$ then $\tau=\sigma-1$ and $I'=\hf$.  We now have
\beqa
\rcg{(1,1)}{0;\frac{1}{2}}{(\lambda,0)}{a_1}{(\lambda,0)}{a_1-1}&=&\sqrt{\frac{3a_1(\lambda-a_1+1)}{2\lambda (\lambda+3)}} \, .\nonumber
\eeqa
and a few straightforward steps yield
\beqa
&&U_{\hbox{\tiny su(3)}}\left((1,1),(\lambda,0),(\sigma-1,\sigma-1),(0,\lambda);(\lambda,0),(\sigma,\sigma)\right)\nonumber \\
&&\qquad =\frac{(\sigma+1)}{2}\sqrt{\frac{3(\lambda-\sigma+1)(\lambda+\sigma+2)}{\lambda(\lambda+3)\sigma(2\sigma+1)}}\, .
\eeqa


\section{Calculation of $\ac^L$ and $\ac^R$ coefficients and their asymptotics}\label{acoefficients}

The  coefficients $\ac^L_{\nu_1 I}(\lambda;\tau)$ have the following general form:
\beqa
&&\ac^L_{\nu I}(\lambda;\tau) =\sqrt{\frac{16}{(\lambda+1)(\lambda+2)}}
\left[\sum_{\sigma=\tau-1}^{\tau+1}\frac{F^\sigma_\lambda}{F^\tau_\lambda} \times
 \right. \nonumber \\
&&\left. 
\sum_{\rho} \rcg{(1,1)}{\nu_1 I}{(\sigma,\sigma)}{\sigma;0}{(\tau,\tau)}{\bar\nu_1;I}_\rho 
U_{\rho}\left[(1,1)(\lambda,0)(\tau,\tau)(0,\lambda);(\lambda,0)(\sigma,\sigma)\right] 
\right]
\label{a}
\eeqa
where the required four Racah coefficient  are given in Table \ref{tableofRacahU}.  
Using the appropriate $SU(3)$ CG coefficients one obtains the explicit expressions 
\beqa
&N \ac^L_{2 \frac{1}{2}}(\lambda;\tau)=\sqrt{3\tau(\tau+2)}  
\left(\frac{\tau  \sqrt{(\lambda -\tau +1) (\lambda +\tau +2)}}{(\tau +1) (2 \tau +1)}\right.
\nonumber \\
&\qquad \qquad 
\left. -\frac{(\tau +2) \sqrt{\lambda -\tau } \sqrt{\lambda +\tau +3}}{(\tau +1) (2\tau +3)}  
+\frac{2 (\lambda -2 \tau  (\tau +2))}{4 \tau  (\tau +2)+3}\right)\, , \\
&N \ac^L_{11}(\lambda;\tau)= \frac{\tau(\tau+2)}{(\tau+1) (2 \tau+1) (
   2 \tau+3) }\left(-2 (2 \lambda +3) (\tau +1) \right.\nonumber \\
&\quad \left. +\sqrt{(\lambda -\tau ) (\lambda +\tau +3)}
+(2 \tau +3) \sqrt{(\lambda -\tau +1) (\lambda +\tau +2)}\right) , \\
&N \ac^L_{10}(\lambda;\tau)= 
\frac{3 \tau^2 \sqrt{(\lambda -\tau +1) (\lambda +\tau +2)}}
{(\tau +1) (2 \tau +1)} \nonumber \\
& \qquad \quad  +\frac{2 (2 \lambda +3) (\tau +2) \tau }{4 \tau  (\tau +2)+3}
+\frac{3(\tau +2)^2 \sqrt{\lambda -\tau } \sqrt{\lambda +\tau +3}}{(\tau +1) (2 \tau +3)}\, ,\\
&N \ac^L_{0 \frac{1}{2}}(\lambda;\tau-1) =\sqrt{3\tau(\tau+2)}
\left(
\frac{\tau  \sqrt{(\lambda -\tau +1) (\lambda +\tau +2)}}{(\tau +1) (2 \tau +1)} \right. \nonumber \\
&\qquad  
\left. 
-\frac{ (\tau +2) \sqrt{\lambda -\tau } \sqrt{\lambda +\tau
   +3}}{(\tau +1) (2 \tau +3)}+\frac{2  (\lambda +2 \tau  (\tau +2)+3)}{4 \tau  (\tau +2)+3}
   \right)\,, 
\eeqa
where the normalization factor $N$ is given in (\ref{N}).
The coefficients $\ac^R_{\nu_1 I}(\lambda;\tau)$ are evaluated in the same manner,
yielding Eq.(\ref{br2half}).

In the limit of large dimension of a representation
the coefficients  $\ac^{L,R}_{\tau; \nu_1 }(\lambda;\bar \nu_1, I)$ can be expanded in inverse powers of semiclassical parameter $\epsilon$:
\beqa
N \ac^L_{11}(\lambda;\tau)=N \ac^R_{11};(\lambda;\tau) \sim -\epsilon \tau(\tau+2)\, ,\\
N \ac^L_{ 10}(\lambda;\tau)=N \ac^R_{10}(\lambda;\tau,0) \sim \frac{2}{\epsilon} - 
3 \epsilon (\tau  (\tau +2)+3)\, ,\\
N \ac^L_{0\frac{1}{2}}(\lambda;\tau)=
N \ac^R_{0\frac{1}{2}}(\lambda;\tau) \sim \sqrt{3\tau(\tau+2)} \left(1+\frac{3 \epsilon}{2}\right) ,\\
N \ac^L_{2\frac{1}{2}}(\lambda;\tau)=N \ac^R_{0\frac{1}{2}}(\lambda;\tau)\sim \sqrt{3\tau(\tau+2)} \left(-1+\frac{3 \epsilon}{2}\right).
\eeqa

\section{Differential operators} 

\subsection{The Casimir operator}\label{sec:differentialCasimir}

\beqa 
&&\hat{\cal C}_2=
-2 \frac{\partial^2}{\partial \beta_2^2}
-\frac{4 }{1-\cos\left(\beta _2\right)}
\frac{\partial^2}{\partial \beta_1^2} \nonumber 
\\
&&+\frac{1}{2}  
  \left(\cos \left(\beta _1\right) \left(\cos \left(\beta _2\right)-1\right)
 -\cos \left(\beta _2\right)-3\right) \csc^2\left(\beta _2\right) \sec ^2\left(\frac{\beta _1}{2}\right) 
 \frac{\partial^2}{\partial \alpha_2^2} \nonumber \\
 &&- 2 \left(2 \cot \left(\beta _2\right)+\csc \left(\beta _2\right)\right) 
 \frac{\partial}{\partial \beta_2}
 +\csc ^2\left(\frac{\beta _2}{2}\right) \sec ^2\left(\frac{\beta _1}{2}\right) 
\frac{\partial^2}{\partial \alpha_1\partial\alpha_2} \nonumber \\
&&+ 2 \csc ^2\left(\beta _1\right) \csc ^2\left(\frac{\beta _2}{2}\right) 
 \frac{\partial^2}{\partial \alpha_1^2}
 -2 \cot \left(\beta _1\right) \csc ^2\left(\frac{\beta _2}{2}\right) 
 \frac{\partial}{\partial \beta_1}
\eeqa 



\subsection{Anaturdjacency relations}\label{adjacencyrelations}

We want to replace the functions $D^{(\tau,\tau)}_{\mu j;\bar\nu I}(\Omega)$ 
with differential operators $\wz_{\nu I}$ acting on functions of the type $D^{(\tau,\tau)}_{\mu j;(\tau\tau\tau)0}(\Omega)$, {\it i.e.} 
we need to find differential operators $\wz_{\nu I}$ acting on these functions so that
\beqa
\wz_{\nu I}D^{(\tau,\tau)}_{\mu J;(\tau\tau\tau)0}(\Omega)
\propto D^{(\tau,\tau)}_{\mu J;\bar\nu I}(\Omega)\, .
\eeqa

First we can recast this as follows.  Let $\Omega_k\in \{\alpha_1,\beta_1,\alpha_2,\beta_2\}$ and start with
\beqa
&&\frac{\partial}{\partial \Omega_k}
D^{(\tau,\tau)}_{\mu J;(\tau,\tau\tau)0}(\Omega)=
\frac{\partial}{\partial \Omega_k}\langle (\tau,\tau)\mu J \vert
\hat{R}(\Omega)\vert (\tau,\tau) \tau\tau\tau;0\rangle\, \nonumber\\
&&\qquad\qquad = \langle (\tau,\tau)\mu J \vert
\frac{\partial}{\partial \Omega_k}\hat{R}(\Omega)\vert  (\tau,\tau) \tau\tau\tau; 0\rangle\, \label{basicdiffaction}\\
&&\qquad\qquad= \langle (\tau,\tau)\mu J \vert
\hat{R}(\Omega)\left(\sum_{\nu I}c_{\nu I}(\Omega_k)\Cop_{\nu I}\right)\vert (\tau,\tau)  \tau\tau\tau; 0\rangle\, ,
\eeqa
where Table \ref{tableofCops} gives the $\Cop_{\nu I}$ in terms of the $\hat{C}_{ij}$.

\begin{table}[H]\centering 
{\renewcommand{\arraystretch}{1.6}
\caption{The relation between $\Cop_{\nu I}$ and generators}
\label{tableofCops}
\begin{tabular}{| C | L || C | L |}
\hline 
\Cop_{210;\frac{1}{2}}& \hat{C}_{13} & \Cop_{201;\frac{1}{2}}& -\hat{C}_{12} \\ \hline 
\Cop_{120;1} & \hat{C}_{23} & \Cop_{111;1}&-\frac{1}{\sqrt{2}}\left(\hat{C}_{22}-\hat{C}_{33}\right) \\ \hline 
\Cop_{102;1} & -\hat{C}_{32} &  \Cop_{111;0}&\frac{1}{\sqrt{6}}\left(2\hat{C}_{11}-\hat{C}_{22}-C_{33}\right)\\ \hline 
\Cop_{021;\frac{1}{2}}& \hat{C}_{21} & \Cop_{012;\frac{1}{2}}& \hat{C}_{31} \\
\hline
\end{tabular}
}
\end{table}

Defined in this way, the operators $\Cop_{\nu I}$ differ from the generators $\hat{C}_{ij}$ by at most a sign and from the tensor 
operators 
$\hat T^{\lambda}_{1;\nu I}$ by a normalization that is a function of the $\mathfrak{su}(3)$ quadratic Casimir invariant and the dimension 
of the irrep on which  the tensors act.

From this we now have the general relation
\beqa
\frac{\partial}{\partial \Omega_k}\hat{R}(\Omega)&= \sum_{\nu I}c_{\nu I}(\Omega_k)\hat{R}(\Omega) \Cop_{\nu I} \label{eq:ccoefficients}
\eeqa

It is important to notice that this relation does not depend on the $\mathfrak{su}(3)$ irrep so the coefficients 
$c_{\nu I}(\Omega_k)$ can be found using any irrep.  The most expeditious
choice is the $3\times 3$ irrep $(1,0)$.  For this representation the operators $\Cop_{\nu I}$ are orthonormal under trace:
\beq
\hbox{Tr}\left( (\Cop_{\nu' I'})^\dagger \Cop_{\nu I}\right)=\delta_{\nu'\nu}\delta_{I'I} \qquad\qquad \qquad  
\eeq
so we can easily write
\beq
c_{\nu' I'}(\Omega_k) =\hbox{Tr}\left((\Cop_{\nu' I'})^\dagger \hat{R}^\dagger(\Omega)\frac{\partial}{\partial \Omega_k}\hat{R}(\Omega)\right) \, .
\eeq

The coefficients $c_{\nu I}(\Omega_k)$ are given in Table \ref{tableofccoefficients}.

\begin{table}[H] \centering
\caption{The $c_{\nu I}(\Omega_k)$ coefficients of equation (\ref{eq:ccoefficients}).}\label{tableofccoefficients}
{\renewcommand{\arraystretch}{1.6}
\begin{tabular}{| C | C | C |}
\hhline{|=|=|=|}
{\nu I} & c_{\nu I}(\alpha_1) & c_{\nu I}(\beta_1)  \\
\hline
210;\frac{1}{2} & i e^{-i \left(\alpha _1+\alpha _2\right)} 
 \sin \left(\frac{\beta _1}{2}\right) \left(\cos ^2\left(\frac{\beta_2}{4}\right)
 +\cos \left(\beta _1\right) \sin ^2\left(\frac{\beta _2}{4}\right)\right) 
 \sin \left(\frac{\beta _2}{2}\right) 
 & -\frac{1}{2} e^{-i \left(\alpha_1+\alpha _2\right)} \cos \left(\frac{\beta _1}{2}\right) 
 \sin \left(\frac{\beta _2}{2}\right) \\
201;\frac{1}{2}
 & i  e^{-i \alpha _2} \sin \left(\frac{\beta _1}{2}\right) \sin \left(\beta_1\right) \sin ^2\left(\frac{\beta _2}{4}\right) 
 \sin \left(\frac{\beta _2}{2}\right) 
 & -\frac{1}{2} e^{-i \alpha _2} \sin \left(\frac{\beta _1}{2}\right) \sin\left(\frac{\beta _2}{2}\right) \\
120;1
 & -i e^{-i \alpha _1} \sin \left(\beta _1\right) \sin ^2\left(\frac{\beta _2}{4}\right) \left(\cos ^2\left(\frac{\beta _2}{4}\right)+\cos \left(\beta
   _1\right) \sin ^2\left(\frac{\beta _2}{4}\right)\right) & e^{-i \alpha _1} \sin ^2\left(\frac{\beta _2}{4}\right) \\   
111;1 & -2 i \sqrt{2} \sin ^2\left(\frac{\beta _1}{2}\right) \sin ^2\left(\frac{\beta _2}{4}\right) \left(\cos \left(\beta _1\right) \sin
   ^2\left(\frac{\beta _2}{4}\right)+1\right) & 0 \\
102;1 & i e^{i \alpha _1} \sin \left(\beta _1\right) \sin ^2\left(\frac{\beta _2}{4}\right) 
 \left(\cos ^2\left(\frac{\beta _2}{4}\right)+\cos \left(\beta_1\right) 
 \sin ^2\left(\frac{\beta _2}{4}\right)\right) & e^{i \alpha _1} \sin ^2\left(\frac{\beta _2}{4}\right) \\
111;0 & 
i \sqrt{\frac{3}{2}} \sin ^2\left(\frac{\beta _1}{2}\right) \sin ^2\left(\frac{\beta _2}{2}\right) & 0 \\
 021;\frac{1}{2} & -i e^{i \alpha _2}\sin \left(\frac{\beta _1}{2}\right) \sin \left(\beta
   _1\right) \sin ^2\left(\frac{\beta _2}{4}\right) \sin \left(\frac{\beta _2}{2}\right) & -\frac{1}{2} e^{i \alpha _2} \sin \left(\frac{\beta _1}{2}\right) \sin
   \left(\frac{\beta _2}{2}\right) \\
    012;\frac{1}{2} & i e^{i \left(\alpha _1+\alpha _2\right)} \sin \left(\frac{\beta _1}{2}\right) \left(\cos ^2\left(\frac{\beta
   _2}{4}\right)+\cos \left(\beta _1\right) \sin ^2\left(\frac{\beta _2}{4}\right)\right) \sin \left(\frac{\beta _2}{2}\right) & \frac{1}{2} e^{i \left(\alpha
   _1+\alpha _2\right)} \cos \left(\frac{\beta _1}{2}\right) \sin \left(\frac{\beta _2}{2}\right) \\
\hhline{|=|=|=|}
{\nu I} & c_{\nu I}(\alpha_2) & c_{\nu I}(\beta_2)  \\  \hline
 210;\frac{1}{2} & \frac{1}{2} i e^{-i \left(\alpha _1+\alpha _2\right)} \sin \left(\frac{\beta _1}{2}\right) \sin \left(\beta _2\right) &
   -\frac{1}{2} e^{-i \left(\alpha _1+\alpha _2\right)} \sin \left(\frac{\beta _1}{2}\right) \\
 201\frac{1}{2} & -\frac{1}{2} i e^{-i \alpha _2} \cos \left(\frac{\beta _1}{2}\right) \sin \left(\beta _2\right) & \frac{1}{2} e^{-i \alpha _2}
   \cos \left(\frac{\beta _1}{2}\right) \\
120;1 & -\frac{1}{2} i e^{-i \alpha _1} \sin \left(\beta _1\right) \sin ^2\left(\frac{\beta _2}{2}\right) & 0 \\
111;1 & \frac{i \cos \left(\beta _1\right) \sin ^2\left(\frac{\beta _2}{2}\right)}{\sqrt{2}} & 0 \\
102;1 & \frac{1}{2} i e^{i \alpha _1} \sin \left(\beta _1\right) \sin ^2\left(\frac{\beta _2}{2}\right) & 0 \\
111;0 & i \sqrt{\frac{3}{2}} \sin ^2\left(\frac{\beta _2}{2}\right) & 0 \\
021;\frac{1}{2} & \frac{1}{2} i e^{i \alpha _2} \cos \left(\frac{\beta _1}{2}\right) \sin \left(\beta _2\right) & \frac{1}{2} e^{i \alpha _2}
   \cos \left(\frac{\beta _1}{2}\right) \\
012;\frac{1}{2} & \frac{1}{2} i e^{i \left(\alpha _1+\alpha _2\right)} \sin \left(\frac{\beta _1}{2}\right) \sin \left(\beta _2\right) &
   \frac{1}{2} e^{i \left(\alpha _1+\alpha _2\right)} \sin \left(\frac{\beta _1}{2}\right) \\ \hline
\end{tabular}
}
\end{table}

To continue, it is convenient to divide the generators in two sets.  The first contains elements in the 
$\mathfrak{u}(2)$ subalgebra: 
$\{\Cop_{120;1},\Cop_{111;1},\Cop_{102;1},\Cop_{111;0}\}$ and will be labeled by roman letters $a,b,c\ldots $. 
The second contains the remaining operators 
$\{\Cop_{210;\frac{1}{2}},\Cop_{201;\frac{1}{2}}\Cop_{021;\frac{1}{2}},\Cop_{012;\frac{1}{2}}\}$ and will be labeled using 
Greek letters $\alpha,\beta\ldots$.

Consider now
\beqa
&&\sum_{k}d_{\beta}(\Omega_k)\frac{\partial }{\partial \Omega_k} \hat{R}(\Omega)\nonumber \\
&&\quad =
\sum_{k\alpha }d_{\beta} (\Omega_k) c_{\alpha}(\Omega_k)\hat{R}(\Omega) \Cop_\alpha + 
\sum_{ak }d_{\beta}(\Omega_k)c_{a}(\Omega_k)\hat{R}(\Omega) \Cop_a
\label{badandgoodguys}
\eeqa
and choose $d_{\beta}(\Omega_k)$ so that 
\beqa
\sum_{k}d_{\beta}(\Omega_k)c_{\alpha}(\Omega_k)&=\delta_{\beta\alpha}\, , 
\label{dcoeff1}
\eeqa
so yielding
\beqa
\sum_{k}d_{\beta}(\Omega_k)\frac{\partial }{\partial \Omega_k} \hat{R}(\Omega)&=
\hat{R}(\Omega) \Cop_\beta + 
\sum_{ak} d_{\beta}(\Omega_k)c_{a}(\Omega_k)  \hat{R}(\Omega) \Cop_a\, . 
\label{dcoeff2}
\eeqa
If we recall from Eq.(\ref{basicdiffaction}) that this sum will act on $\ket{(\tau,\tau)\tau\tau\tau;0}$, and that
 $\ket{(\tau,\tau)\tau\tau\tau;0}$ is by construction annihilated by $\Cop_a$, we see that Eq.(\ref{dcoeff1}) is simply a linear system for $d_\beta$ which 
can be easily solved.
The solution coefficients  $d_{\beta}(\Omega_k)$ are found in Table \ref{tableofdcoefficients}.

\begin{table}[H] 
\caption{The $d_{\nu I}(\Omega_k)$ coefficients.}\label{tableofdcoefficients}
{\renewcommand{\arraystretch}{1.6}
\hskip-2.0cm
\begin{tabular}{| C | C | C |}
\hline
&{\nu I}=210;\textstyle\frac{1}{2}&{\nu I}=201;\textstyle\frac{1}{2}\\ 
\hline
d_{\nu I}(\alpha_1) & 
 -\frac{1}{2} i e^{i \left(\alpha _1+\alpha _2\right)} \csc \left(\frac{\beta _1}{2}\right) 
 \csc \left(\frac{\beta _2}{2}\right) & 
 -\frac{i}{2} e^{i \alpha _2} \csc \left(\frac{\beta_2}{2}\right) 
 \sec \left(\frac{\beta _1}{2}\right) \\
 d_{\nu I}(\beta_1) &
 -e^{i \left(\alpha _1+\alpha _2\right)} \cos \left(\frac{\beta _1}{2}\right) \csc \left(\frac{\beta _2}{2}\right) & -e^{i \alpha _2} \sin \left(\frac{\beta _1}{2}\right) 
 \csc \left(\frac{\beta_2}{2}\right) \\
d_{\nu I}(\alpha_2)&   
 -2 i e^{i \left(\alpha _1+\alpha _2\right)} \sin \left(\frac{\beta _1}{2}\right) \sin ^2\left(\frac{\beta _2}{4}\right) \csc \left(\beta _2\right) & 
 \mbox{$\begin{array}{ll}-\frac{i}{2} \sin \left(\frac{\beta_1}{2}\right) 
   e^{i\alpha_2}  \left(\cot \left(\beta _1\right) \csc \left(\frac{\beta _2}{2}\right) \right.\\ 
   \left. \qquad -2 \left(\cos \left(\beta _1\right)+
   \cos\left(\frac{\beta _2}{2}\right)+1\right) \csc \left(\beta _1\right) \csc \left(\beta _2\right)\right)
   \end{array}$}\\
d_{\nu I}(\beta_2)&
 -e^{i \left(\alpha _1+\alpha _2\right)} \sin \left(\frac{\beta _1}{2}\right) & 
 e^{i \alpha _2} \cos \left(\frac{\beta _1}{2}\right)  \\
 \hhline{|=|=|=|}
 &{\nu I}=012;\textstyle\frac{1}{2}&{\nu I}=021;\textstyle\frac{1}{2}\\ 
\hline
d_{\nu I}(\alpha_1) & 
-\frac{1}{2} i e^{-i \left(\alpha _1+\alpha _2\right)} \csc \left(\frac{\beta _1}{2}\right) \csc \left(\frac{\beta _2}{2}\right)& 
\frac{i}{2}e^{-i\alpha_2} \csc \left(\frac{\beta _2}{2}\right) \sec \left(\frac{\beta _1}{2}\right) 
\\
 d_{\nu I}(\beta_1) &
e^{-i \left(\alpha _1+\alpha _2\right)} \cos \left(\frac{\beta _1}{2}\right) \csc \left(\frac{\beta _2}{2}\right)
& -e^{-i \alpha _2} \sin \left(\frac{\beta _1}{2}\right) \csc \left(\frac{\beta _2}{2}\right)
 \csc \left(\frac{\beta_2}{2}\right) \\
d_{\nu I}(\alpha_2)&   
-2 i e^{-i \left(\alpha _1+\alpha _2\right)} \sin \left(\frac{\beta _1}{2}\right) \sin ^2\left(\frac{\beta _2}{4}\right) \csc \left(\beta _2\right)& 
 \mbox{$\begin{array}{ll}\frac{1}{2} i e^{-i \alpha _2} \sin \left(\frac{\beta _1}{2}\right) \left(\cot \left(\beta _1\right) \csc \left(\frac{\beta _2}{2}\right)
 \right. \\
  \left. \qquad -2 \left(\cos
   \left(\beta _1\right)+\cos \left(\frac{\beta _2}{2}\right)+1\right) \csc \left(\beta _1\right) \csc \left(\beta _2\right)\right)
   \end{array}$}
 \\
d_{\nu I}(\beta_2)&
e^{-i \left(\alpha _1+\alpha _2\right)} \sin \left(\frac{\beta _1}{2}\right)& 
e^{-i \alpha _2} \cos \left(\frac{\beta _1}{2}\right) \\
\hline
\end{tabular}
}
\end{table}

With this:
\beqa
\wz_{\nu \frac{1}{2}} D^{(\tau,\tau)}_{\mu J,(\tau\tau\tau)0}(\Omega)&= 
\sum_{k}d_{\nu \frac{1}{2}}(\Omega_k)
\frac{\partial }{\partial \Omega_k} D^{(\tau,\tau)}_{\mu J,(\tau\tau\tau)0}(\Omega)\, \label{eq:differentialactionraw}
\eeqa
where $\wz_{\nu \frac{1}{2}}$ is a differential operator that shifts the function $D^{(\tau,\tau)}_{\mu J,(\tau\tau\tau)0}(\Omega)$ to 
$D^{(\tau,\tau)}_{\mu J; \bar{\nu} \frac{1}{2}}(\Omega)$ up to a proportionality term. 

If we substitute Eq.(\ref{badandgoodguys}) into Eq.(\ref{eq:differentialactionraw}) we find
\beqa
&&\wz_{\nu \frac{1}{2}} D^{(\tau,\tau)}_{\mu J,(\tau\tau\tau)0}(\Omega)\nonumber \\
&&\qquad =\langle (\tau,\tau)\mu J\vert \hat{R}(\Omega) \left(\Cop_{\nu \frac{1}{2}} + 
\sum_{ak }d_{\nu\frac{1}{2}}(\Omega_k)c_{a}(\Omega_k) \Cop_a\right)\vert(\tau,\tau)\tau\tau\tau;0\rangle\, \nonumber \\
&&\qquad=\langle (\tau,\tau)\mu J\vert \hat{R}(\Omega) \Cop_{\nu \frac{1}{2}} \vert(\tau,\tau)\tau\tau\tau;0\rangle\, \nonumber\\
&&\qquad= \langle (\tau,\tau)\mu J\vert \hat{R}(\Omega) \vert(\tau,\tau)\bar\nu ;
\textstyle\frac{1}{2}\rangle\langle (\tau,\tau) \bar\nu;\textstyle\frac{1}{2}\vert \Cop_{\nu \frac{1}{2}} \vert(\tau,\tau)\tau\tau\tau;0\rangle\, \nonumber\\
&&\qquad=D^{(\tau,\tau)}_{\mu J; \bar\nu \frac{1}{2}}(\Omega)
\langle (\tau,\tau) \bar\nu ;\textstyle\frac{1}{2}\vert \Cop_{\nu \frac{1}{2}} \vert(\tau,\tau)\tau\tau\tau;0\rangle
\label{eq:Atildeaction}
\eeqa
since $\Cop_a\vert(\tau,\tau)\tau\tau\tau;0\rangle=0$.

Finally, we can evaluate $\langle (\tau,\tau) \bar\nu ;\textstyle\frac{1}{2}\vert \Cop_{\nu \frac{1}{2}} \vert(\tau,\tau)\tau\tau\tau;0\rangle$.  
It turns out that this expression is quite simply expressed in terms of $\nu$: 
\beqa
\langle (\tau,\tau) \bar\nu ;\textstyle\frac{1}{2}\vert \Cop_{\nu \frac{1}{2}} \vert(\tau,\tau)\tau\tau\tau;0\rangle
=(-1)^{\bar\nu_1/2}\sqrt{\frac{\tau(\tau+2)}{2}}\, ,
\eeqa
where $\nu$ and $\bar\nu$ are related in Table 
\ref{table:nunubar}. Combining this with Eq.(\ref{eq:Atildeaction}), we now have
\beqa
\wz_{\nu \frac{1}{2}} 
D^{(\tau,\tau)}_{\mu J;(\tau\tau\tau)0}(\Omega)
&=
(-1)^{\bar\nu_1/2}\sqrt{\frac{\tau(\tau+2)}{2}}D^{(\tau,\tau)}_{\mu J;\bar\nu \frac{1}{2}}(\Omega) \,  \label{eq:shiftonehalf} \\
&=\langle {(\tau,\tau)\mu J}
\vert R(\Omega)\Cop_{\nu;\frac{1}{2}}\vert (\tau,\tau)\tau\tau\tau;0\rangle \, \quad ,
 \label{eq:tinsideshiftonehalf}
\eeqa
where the $\wz_{\nu \frac{1}{2}}$ operators are of first order only.


Next we investigate  adjacency relations of the form 
\beq 
\wz^{(2)}_{(1\nu_2\nu_3) I}D^{(\sigma,\sigma)}_{\mu J;\bar\nu 1}(\Omega)\, ,
\eeq
and show, in agreement with the argument presented in Sec.\ref{sec:rules}, that this operator is of second order in the derivatives.

We consider
\beqa
&&\wz_{\alpha}\wz_{\beta}D^{(\sigma,\sigma)}_{\mu J;(\sigma\sigma\sigma)0}(\Omega)\nonumber \\
&&=\langle (\sigma,\sigma)\mu J\vert \hat{R}(\Omega)
\left(\Cop_{\alpha}+\sum_{ak }d_{\alpha}(\Omega_k)c_{a}(\Omega_k) \Cop_a\right)\Cop_{\beta}\vert(\sigma,\sigma)\sigma\sigma\sigma;0\rangle \, ,
\nonumber \\
&&=\langle (\sigma,\sigma)\mu J\vert \hat{R}(\Omega)
\Cop_{\alpha}\Cop_{\beta}\vert(\sigma,\sigma)\sigma\sigma\sigma;0\rangle \nonumber \\
&&\quad +\langle (\sigma,\sigma)\mu J\vert \hat{R}(\Omega)
\sum_{ak }d_{\beta}(\Omega_k)c_{a}(\Omega_k) \Cop_a \Cop_{\beta}\vert(\sigma,\sigma)\sigma\sigma\sigma;0\rangle \, .
\eeqa
Now, since $\Cop_{\beta}$ is an element of the $\mathfrak{u}(2)$ subalgebra, we have
\beqa
\Cop_a \Cop_{\beta} \vert(\sigma,\sigma)\sigma\sigma\sigma;0\rangle&= 
[\Cop_a,\Cop_\beta]\vert(\sigma,\sigma)\sigma\sigma\sigma;0\rangle+ \hat{C}_\beta \hat{C}_a\vert(\sigma,\sigma)\sigma\sigma\sigma;0\rangle\, \nonumber\\
&=[\Cop_a,\Cop_\beta]\vert(\sigma,\sigma)\sigma\sigma\sigma;0\rangle\, \nonumber\\
&=\sum_{\gamma}g_{a\beta}^\gamma \Cop_\gamma\vert(\sigma,\sigma)\sigma\sigma\sigma;0\rangle
\eeqa
As we have for $\Cop_\gamma$
\beqa
\wz_\gamma D^{(\sigma,\sigma)}_{\mu J;(\sigma\sigma\sigma)0}(\Omega)=
\langle (\sigma,\sigma)\mu J\vert \hat{R}(\Omega) \Cop_\gamma \vert(\sigma,\sigma)\sigma\sigma\sigma;0\rangle\, ,
\eeqa 
we find that 
\beqa
&&\wz_{\alpha}\wz_{\beta}D^{(\sigma,\sigma)}_{\mu J;(\sigma\sigma\sigma)0}(\Omega)\nonumber \\
&&=\langle (\sigma,\sigma)\mu J\vert \hat{R}(\Omega)
\left(\Cop_{\alpha}+\sum_{ak }d_{\alpha}(\Omega_k)c_{a}(\Omega_k) \Cop_a\right)\Cop_{\beta}\vert(\sigma,\sigma)\sigma\sigma\sigma;0\rangle \, \nonumber\\
&&=\langle (\sigma,\sigma)\mu J\vert \hat{R}(\Omega)
\Cop_{\alpha}\Cop_{\beta}\vert(\sigma,\sigma)\sigma\sigma\sigma;0\rangle\, \nonumber\\
&&\qquad +\sum_{ak\gamma }d_{\alpha}(\Omega_k)c_{a}(\Omega_k)g_{a\beta}^\gamma
\wz_\gamma D^{(\sigma,\sigma)}_{\mu J;(\sigma\sigma\sigma)0}(\Omega) \, 
\eeqa
or
\beqa
&&\left(\wz_{\alpha}\wz_{\beta}
-\sum_{a\gamma }f_{\alpha a} g_{a\beta}^\gamma
\wz_\gamma \right) D^{(\sigma,\sigma)}_{\mu J;(\sigma\sigma\sigma)0}(\Omega)  \nonumber \\
&&=\langle (\sigma,\sigma)\mu J\vert \hat{R}(\Omega)
\Cop_{\alpha}\Cop_{\beta}\vert(\sigma,\sigma)\sigma\sigma\sigma;0\rangle\, , \label{shiftby1}
\eeqa
where, for economy, we denote
\beqa
f_{\alpha a}:&= \sum_{k}d_{\alpha}(\Omega_k)c_a(\Omega_k)\, .
\label{faacoeff}
\eeqa
These coefficients are given in Table  \ref{tableofsumsdc}.

\begin{table}[h] \centering
\caption{The coefficients $f_{\beta a}$.}\label{tableofsumsdc}
{\renewcommand{\arraystretch}{1.6}
\hskip-2.25cm\begin{tabular}{| C | C | C |}
\hline
\beta & a = 120;1
& 111;1  
\\ \hline
012;\frac{1}{2} 
&
\begin{array}{l}
-2 e^{-i \left(2 \alpha _1+\alpha _2\right)} \sin \left(\frac{\beta _1}{2}\right) \\
 \times \sin \left(\beta _1\right) \sin ^4\left(\frac{\beta _2}{4}\right) \csc \left(\beta _2\right) 
 \end{array}
 & 
 \begin{array}{l}
 e^{-i \left(\alpha _1+\alpha _2\right)} \sin \left(\frac{\beta _1}{2}\right) 
 \tan \left(\frac{\beta _2}{4}\right)  \\
\times \frac{1}{\sqrt{2}}\left(\sin ^2\left(\frac{\beta _2}{4}\right) \cos \left(\beta _1\right) \sec
   \left(\frac{\beta _2}{2}\right)-1\right)\end{array} 
\\
\hline 
021;\frac{1}{2}
&
\begin{array}{l}
-\frac{1}{4} e^{-i \left(\alpha _1+\alpha _2\right)} \sin \left(\frac{\beta _1}{2}\right) 
\tan \left(\frac{\beta _2}{4}\right)  \\ 
\times \left(\sec \left(\frac{\beta _2}{2}\right) \left(2 \sin^2\left(\frac{\beta _2}{4}\right) 
\cos \left(\beta _1\right)+1\right)+3\right)
 \end{array}
 & 
 \begin{array}{l}
\frac{1}{\sqrt{2}}
e^{-i \alpha _2} \cos \left(\frac{\beta _1}{2}\right) \tan \left(\frac{\beta _2}{4}\right)
 \\ 
\times \left(\sin ^2\left(\frac{\beta _2}{4}\right) \cos \left(\beta _1\right) 
\sec \left(\frac{\beta_2}{2}\right)+1\right)\end{array}\\
\hline
201;\frac{1}{2} & 
\begin{array}{l}
e^{-i \alpha_1} \sin \alpha_2 (\cot\alpha_2+i) \\ 
\times \sin ^2\beta_1 \csc \left(\frac{\beta_1}{2}\right) \sin ^4\left(\frac{\beta_2}{4}\right) \csc \beta_2
\end{array} & 
\begin{array}{l}
-\frac{1}{\sqrt{2}}e^{i \alpha _2} \cos \left(\frac{\beta _1}{2}\right) \tan \left(\frac{\beta _2}{4}\right) \\
\times \left(\sin ^2\left(\frac{\beta _2}{4}\right) \cos \beta_1 \sec \left(\frac{\beta
   _2}{2}\right)+1\right)
\end{array} \\
\hline
210;\frac{1}{2} & \begin{array}{l}
\qquad e^{i \alpha _2} \cos \left(\frac{\beta _1}{2}\right) \tan \left(\frac{\beta _2}{4}\right) \\
\times \frac{1}{4}\left(\sec \left(\frac{\beta _2}{2}\right) \left(2 \sin ^2\left(\frac{\beta _2}{4}\right) \cos\beta_1-1\right)-3\right)
\end{array}
& \begin{array}{l}
e^{i \left(\alpha _1+\alpha _2\right)} \sin \left(\frac{\beta _1}{2}\right) \tan \left(\frac{\beta _2}{4}\right) \\
\times \frac{1}{\sqrt{2}}\left(\sin ^2\left(\frac{\beta _2}{4}\right) \cos\beta_1 \sec \left(\frac{\beta
   _2}{2}\right)-1\right)
\end{array} \\
\hline
\beta & a = 102;1
& 111;0 \\
\hline
012;\frac{1}{2}&\begin{array}{l}
\qquad e^{-i \alpha _2} \cos \left(\frac{\beta _1}{2}\right) \tan \left(\frac{\beta _2}{4}\right) \\
\times \frac{1}{4}\left(\sec \left(\frac{\beta _2}{2}\right) \left(1-2 \sin ^2\left(\frac{\beta _2}{4}\right) \cos\beta_1\right)+3\right)
\end{array} & \begin{array}{l}
\frac{1}{2} \sqrt{\frac{3}{2}} e^{-i \left(\alpha _1+\alpha _2\right)} \sin \left(\frac{\beta_1}{2}\right) \tan \left(\frac{\beta_2}{2}\right)
\end{array}\\
\hline
021;\frac{1}{2} & \begin{array}{l}
\qquad -e^{-i \left(\alpha _1+\alpha_2\right)} \sin \left(\frac{\beta _1}{2}\right) \tan \left(\frac{\beta_2}{4}\right) \\
\times \frac{1}{4}\left(\sec \left(\frac{\beta _2}{2}\right) \left(2 \sin ^2\left(\frac{\beta _2}{4}\right) \cos\beta_1+1\right)+3\right) \end{array} & \begin{array}{l}
\frac{1}{2} \sqrt{\frac{3}{2}} e^{-i \alpha_2} \cos \left(\frac{\beta_1}{2}\right) \tan \left(\frac{\beta_2}{2}\right)
\end{array}\\
\hline
201;\frac{1}{2} & \begin{array}{l}
\qquad e^{-i \alpha _1} \sin\alpha_2  \csc \left(\frac{\beta _1}{2}\right) \csc\beta_2 \\
\times  \sin ^2\left(\beta _1\right) \sin ^4\left(\frac{\beta _2}{4}\right)\left(\cot \left(\alpha _2\right)+i\right)
\end{array}  &  \begin{array}{l}
-\frac{1}{2} \sqrt{\frac{3}{2}} e^{i \alpha _2} \cos \left(\frac{\beta _1}{2}\right) \tan \left(\frac{\beta _2}{2}\right)
\end{array}\\
\hline
210;\frac{1}{2}& \begin{array}{l}
\qquad \frac{1}{4} e^{i \alpha_2} \cos \left(\frac{\beta_1}{2}\right) \tan \left(\frac{\beta_2}{4}\right) \\
\times \left(\sec \left(\frac{\beta_2}{2}\right) \left(2 \sin ^2\left(\frac{\beta _2}{4}\right) \cos\beta_1-1\right)-3\right)\end{array} &
\begin{array}{l}
\frac{1}{2} \sqrt{\frac{3}{2}} e^{i \left(\alpha _1+\alpha _2\right)} \sin \left(\frac{\beta _1}{2}\right) \tan \left(\frac{\beta _2}{2}\right)
\end{array} \\
\hline
\end{tabular}
}
\end{table}

\subsection{The Second Order Operator $\wz^{(2)}_{(1\nu_2\nu_3) I}$}

In order to obtain operators
which shifts the $I$ label by $1$, we work with products of the operators $\wz_{\nu\frac{1}{2}}$.

If we consider 
\beqa
&&\wz_{(021)\frac{1}{2}}\wz_{(210)\frac{1}{2}}D^{(\sigma,\sigma)}_{\mu,J;(\sigma,\sigma,\sigma)0}(\Omega) \nonumber \\
&&=\bra{(\sigma,\sigma)\mu;J}\hat{R}(\Omega)
\Cop_{(021)\frac{1}{2}}\Cop_{(210)\frac{1}{2}}\ket{(\sigma,\sigma)\sigma\sigma\sigma;0}\nonumber \\
&&+\sum_{a}f_{(021)\frac{1}{2};a}
\bra{(\sigma,\sigma)\mu;J}\hat{R}(\Omega)
\left[{\Cop_{a}},{\Cop_{(210)\frac{1}{2}}}\right]\ket{(\sigma,\sigma)\sigma\sigma\sigma;0}\, \quad ,
\label{shiftby1specific1}
\eeqa
where the $f_{\alpha a}$ coefficients that appear in Eq.(\ref{shiftby1specific1}) are given in 
Table \ref{tableofsumsdc}. Using
\beqa
&&\sum_{a}f_{(021)\frac{1}{2};a}\bra{(\sigma,\sigma)\mu;J}\hat{R}(\Omega)
\left[\Cop_{a},\Cop_{(210)\frac{1}{2}}\right]
\ket{(\sigma,\sigma)\sigma\sigma\sigma;0}\nonumber \\
&&\quad =
 -f_{(021)\frac{1}{2};(102)\frac{1}{2}}
 \bra{(\sigma,\sigma)\mu;J}\hat{R}(\Omega)\Cop_{(201)\frac{1}{2}}
\ket{(\sigma,\sigma)\sigma\sigma\sigma;0}    \nonumber \\
&&  
\qquad +\left(-\textstyle\frac{\sqrt{2}}{2} f_{(021)\frac{1}{2};(111)1} 
+\textstyle\sqrt{\frac{3}{2}}f_{(021)\frac{1}{2};(111)0}\right)   \\
&&\qquad\qquad\times 
\bra{(\sigma,\sigma)\mu;J} \hat{R}(\Omega)\Cop_{(210)\frac{1}{2}}
\ket{(\sigma,\sigma)\sigma\sigma\sigma;0} \Big)\, \nonumber \\
&&\quad = 
 -f_{(021)\frac{1}{2};(102)\frac{1}{2}}\wz_{(201)\frac{1}{2}}D^{(\sigma,\sigma)}_{\mu J;(\sigma\sigma\sigma)0}(\Omega) 
 \nonumber \\
&&\; +\left(\textstyle\sqrt{\frac{3}{2}}f_{(021)\frac{1}{2};(111)0}-\textstyle\frac{\sqrt{2}}{2} f_{(021)\frac{1}{2};(111)1} 
\right)\wz_{(210)\frac{1}{2}}
D^{(\sigma,\sigma)}_{\mu J;(\sigma\sigma\sigma)0}(\Omega)
\eeqa
and
\beqa
&&\bra{(\sigma,\sigma)\mu;J}\hat{R}(\Omega)\Cop_{(021)\frac{1}{2}}\Cop_{(210)\frac{1}{2}} 
\ket{(\sigma,\sigma)\sigma\sigma\sigma;0}=\nonumber \\
&&\qquad -\frac{\sigma(\sigma+2)}{\sqrt{6}}D^{(\sigma,\sigma)}_{\mu J;(\sigma,\sigma+1,\sigma-1)1}
(\Omega)
\eeqa
we obtain the expression
\beqa
&-\frac{\sigma(\sigma+2)}{\sqrt{6}}D^{(\sigma,\sigma)}_{\mu J;(\sigma,\sigma+1,\sigma-1)1}(\Omega) \nonumber \\
&\quad =\wz_{(021)\frac{1}{2}}\wz_{(210)\frac{1}{2}}D^{(\sigma,\sigma)}_{\mu,J;(\sigma\sigma\sigma)0}\nonumber \\
& +\left(\textstyle\frac{\sqrt{2}}{2}f_{(021)\frac{1}{2};(111)1}\wz_{(210)\frac{1}{2}}
-\textstyle\sqrt{\frac{3}{2}}f_{(021)\frac{1}{2};(111)0}\wz_{(210)\frac{1}{2}} \right)
D^{(\sigma,\sigma)}_{\mu,J;(\sigma\sigma\sigma)0}(\Omega)\, \nonumber \\
&\qquad+ f_{(021)\frac{1}{2};(102)\frac{1}{2}}
\wz_{(201)\frac{1}{2}}D^{(\sigma,\sigma)}_{\mu,J;(\sigma\sigma\sigma)0}(\Omega)
\, ,\\
&\quad:= \wh_{(120);1}
D^{(\sigma,\sigma)}_{\mu,J;(\sigma\sigma\sigma)0}(\Omega) \, . \label{eq:S21201}
\eeqa

Similarly, starting with 
\beqa
&&\bra{(\sigma,\sigma)\mu;J}\hat{R}(\Omega)\Cop_{(012)\frac{1}{2}}\Cop_{(201)\frac{1}{2}}
\ket{(\sigma,\sigma)\sigma\sigma\sigma;0}\nonumber \\
&&=-\frac{\sigma(\sigma+2)}{\sqrt{6}}D^{(\sigma,\sigma)}_{\mu,J;(\sigma,\sigma-1,\sigma+1)1}(\Omega) \,  ,
\eeqa
we easily reach
\beqa
&-\frac{\sigma(\sigma+2)}{\sqrt{6}}D^{(\sigma,\sigma)}_{\mu J;(\sigma,\sigma-1,\sigma+1)1}(\Omega) \nonumber \\
&\quad =\wz_{(012)\frac{1}{2}}\wz_{(201)\frac{1}{2}}D^{(\sigma,\sigma)}_{\mu,J;(\sigma\sigma\sigma)0}\nonumber \\
&+
\left(\textstyle\frac{\sqrt{2}}{2} f_{(012)\frac{1}{2};(111)1}
+\textstyle\sqrt{\frac{3}{2}} f_{(012)\frac{1}{2};(111)0}
\wz_{(201)\frac{1}{2}}\right)D^{(\sigma,\sigma)}_{\mu,J;(\sigma\sigma\sigma)0}(\Omega) \nonumber \\
&\qquad - f_{(012)\frac{1}{2};(120)1}
\wz_{(210)\frac{1}{2}} D^{(\sigma,\sigma)}_{\mu,J;(\sigma\sigma\sigma)0}(\Omega) 
 \\
&\quad:= \wh_{(102);1}D^{(\sigma,\sigma)}_{\mu,J;(\sigma\sigma\sigma)0}(\Omega) \quad .
\label{eq:S21021}
\eeqa

Finally, we consider the action
\beqa
&\bra{(\sigma,\sigma)\mu;J}\hat{R}(\Omega)
\left(\Cop_{021;\frac{1}{2}}\Cop_{201;\frac{1}{2}}+ \Cop_{012;\frac{1}{2}}\Cop_{210;\frac{1}{2}}\right)
\ket{(\sigma,\sigma)\sigma\sigma\sigma;0} 
\nonumber \\
&=-\frac{\sigma(\sigma+2)}{\sqrt{3}}D^{(\sigma,\sigma)}_{\mu J;(\sigma\sigma\sigma)1}(\Omega)\, \nonumber \\
&: = -\frac{1}{\sqrt{3}} \wh_{(111);1}D^{(\sigma,\sigma)}_{\mu,J;(\sigma\sigma\sigma)0}(\Omega)\, .
\label{eq:S21111}
\eeqa
We can then verify that
\beqa
&\wz_{021;\frac{1}{2}}\wz_{201;\frac{1}{2}}D^{(\sigma,\sigma)}_{\mu J;(\sigma\sigma\sigma)}(\Omega)\nonumber \\
&\quad =\bra{(\sigma,\sigma)\mu;J}\hat{R}(\Omega) \Cop_{021;\frac{1}{2}}\Cop_{201;\frac{1}{2}}
\ket{(\sigma,\sigma)\sigma\sigma\sigma;0}\nonumber \\
&+
 f_{(021)\frac{1}{2};(120)1}\bra{(\sigma,\sigma)\mu J}\hat{R}(\Omega)\Cop_{210;\frac{1}{2}}\ket{(\sigma,\sigma)\sigma\sigma\sigma;0}
 \nonumber \\
 &\qquad +
 \left( \textstyle\sqrt{\frac{3}{2}}f_{(021)\frac{1}{2};(111)1} + \textstyle\frac{\sqrt{2}}{2}f_{(021)\frac{1}{2};(111)0}\right) 
 \nonumber \\
&\qquad\qquad \times
\bra{(\sigma,\sigma)\mu J}\hat{R}(\Omega)\Cop_{201;\frac{1}{2}}\ket{(\sigma,\sigma)\sigma\sigma\sigma;0}\,  ,\\
 &\quad =\bra{(\sigma,\sigma)\mu;J}\hat{R}(\Omega) \Cop_{021;\frac{1}{2}}\Cop_{201;\frac{1}{2}}
\ket{(\sigma,\sigma)\sigma\sigma\sigma;0}\nonumber \\
&\qquad + \left( \textstyle\sqrt{\frac{3}{2}}f_{(021)\frac{1}{2};(111)1} + \textstyle\frac{\sqrt{2}}{2}f_{(021)\frac{1}{2};(111)0}\right)
 \wz_{201;\frac{1}{2}} D^{(\sigma,\sigma)}_{\mu J;(\sigma\sigma\sigma)}(\Omega) \nonumber \\
 &\qquad +
f_{(021)\frac{1}{2};(120)1}\wz_{210;\frac{1}{2}}D^{(\sigma,\sigma)}_{\mu J;(\sigma\sigma\sigma)}(\Omega)\, ,
\eeqa
where equation (\ref{eq:tinsideshiftonehalf}) has been used.

Duplicating the same steps, this time for 
$\wz_{012;\frac{1}{2}}\wz_{210;\frac{1}{2}}D^{(\sigma,\sigma)}_{\mu J;(\sigma\sigma\sigma)}(\Omega)$,
yields
\beqa
&\wz_{012;\frac{1}{2}}\wz_{210;\frac{1}{2}}D^{(\sigma,\sigma)}_{\mu J;(\sigma\sigma\sigma)}(\Omega)\nonumber \\
&\quad =\bra{(\sigma,\sigma)\mu;J}\hat{R}(\Omega) \Cop_{012;\frac{1}{2}}\Cop_{210;\frac{1}{2}}
\ket{(\sigma,\sigma)\sigma\sigma\sigma;0}\nonumber \\
&\quad 
-f_{(012)\frac{1}{2};(102)1}\bra{(\sigma,\sigma)\mu J}\hat{R}(\Omega)\Cop_{201;\frac{1}{2}}\ket{(\sigma,\sigma)\sigma\sigma\sigma;0}
 \nonumber \\
 &\qquad +
 \left( \textstyle\sqrt{\frac{3}{2}}f_{(012)\frac{1}{2};(111)1} - \textstyle\frac{\sqrt{2}}{2}f_{(012)\frac{1}{2};(111)0}\right)
 \nonumber \\
&\qquad\qquad  
\times  \bra{(\sigma,\sigma)\mu J}\hat{R}(\Omega)\Cop_{210;\frac{1}{2}}\ket{(\sigma,\sigma)\sigma\sigma\sigma;0}\, , \\
 &\quad =\bra{(\sigma,\sigma)\mu;J}\hat{R}(\Omega) \Cop_{021;\frac{1}{2}}\Cop_{201;\frac{1}{2}}
\ket{(\sigma,\sigma)\sigma\sigma\sigma;0}\nonumber \\
&\qquad 
 + \left( \textstyle\sqrt{\frac{3}{2}}f_{(012)\frac{1}{2};(111)1} - \textstyle\frac{\sqrt{2}}{2}f_{(012)\frac{1}{2};(111)0}\right)
 \wz_{210;\frac{1}{2}} D^{(\sigma,\sigma)}_{\mu J;(\sigma\sigma\sigma)}(\Omega) \nonumber \\
 &\qquad +
 f_{(012)\frac{1}{2};(102)1}\wz_{201;\frac{1}{2}} D^{(\sigma,\sigma)}_{\mu J;(\sigma\sigma\sigma)}(\Omega) \, .
\eeqa
Hence: 
\beqa
& -\frac{1}{\sqrt{3}} \wh_{(111);1}D^{(\sigma,\sigma)}_{\mu,J;(\sigma\sigma\sigma)0}(\Omega)\nonumber \\
&=
\left(\wz_{021;\frac{1}{2}}\wz_{201;\frac{1}{2}}- f_{(021)\frac{1}{2};(120)1}\wz_{210;\frac{1}{2}}\right)
D^{(\sigma,\sigma)}_{\mu J;(\sigma\sigma\sigma)0}(\Omega)\nonumber \\
& - \left( \textstyle\sqrt{\frac{3}{2}}f_{(021)\frac{1}{2};(111)1} + \textstyle\frac{\sqrt{2}}{2}f_{(021)\frac{1}{2};(111)0}\right)
 \wz_{201;\frac{1}{2}} D^{(\sigma,\sigma)}_{\mu J;(\sigma\sigma\sigma)0}(\Omega) \nonumber \\
 &\, + \left(\wz_{012;\frac{1}{2}}\wz_{210;\frac{1}{2}}-
  f_{(012)\frac{1}{2};(102)1}\wz_{201;\frac{1}{2}}\right)D^{(\sigma,\sigma)}_{\mu J;(\sigma\sigma\sigma)0}(\Omega)\nonumber \\
& - \left( \textstyle\sqrt{\frac{3}{2}}f_{(012)\frac{1}{2};(111)1} - \textstyle\frac{\sqrt{2}}{2}f_{(012)\frac{1}{2};(111)0}\right)
 \wz_{210;\frac{1}{2}} D^{(\sigma,\sigma)}_{\mu J;(\sigma\sigma\sigma)0}(\Omega) \, .
 \label{eq:S21111}
\eeqa

We can summarize equations (\ref{eq:S21201}), (\ref{eq:S21021}) and (\ref{eq:S21111}) as
\beqa
&&\wh_{1\nu_2\nu_3;1}D^{(\sigma,\sigma)}_{\mu J;(\sigma\sigma\sigma)0}(\Omega)
\nonumber \\
&&=-
{\sigma(\sigma+2)}\sqrt{\frac{(1+\delta_{\nu_2 1}\delta_{\nu_3 1})}{6}}
D^{(\sigma,\sigma)}_{\mu J;(1\nu_2\nu_3)1}(\Omega)\, .
\label{eq:summaryS2ops}
\eeqa 



\section{Explicit form of Eq.(\ref{H2ex}) for $\alpha J=(111)0$ and $(111)1$}\label{sec:differentialoperators}

\subsection{General expressions}

For $\nu_1'=0,2$, define 
\beqa 
\hat {\cal S}_{\alpha J;
\nu^\prime_1\frac{1}{2}}&:=&
\sum_{\nu^\prime_2\nu^\prime_3}D^{(1,1)}_{(\nu^\prime_1\nu^\prime_2\nu^\prime_3)\frac{1}{2};\alpha J}(\Omega^{-1})
\wz_{(\nu^\prime_1\nu^\prime_2\nu^\prime_3)\frac{1}{2}}\, ,\\
\hat {\cal S}_{\alpha J}&=&
\hat {\cal S}_{\alpha J;0\frac{1}{2}} + 
\hat {\cal S}_{\alpha J;2\frac{1}{2}}
\eeqa 
and denote by $g(\Omega)\equiv g(\alpha_1,\beta_1,\alpha_2,\beta_2)$.  
In this way Eq.(\ref{H2ex}) becomes
\beq 
i\partial_tW_{\hat \rho} (\Omega) =-\frac{\sqrt{24}}{N}\hat{\cal S}_{\alpha J} 
\hat\frak{C}_{\alpha J}W_\rho(\Omega)\, \qquad \hat H=(\hat T^\lambda_{1;\alpha J})^2\, ,
\eeq 
where
\beq 
\hat{\frak C}_{\alpha J}W_{\hat\rho}(\Omega):= (\hat\mathfrak{C}_{\alpha J}^L+\hat\mathfrak{C}_{\alpha J}^R) 
W_{\hat\rho}(\Omega) \eeq 
as per Eq.(\ref{H2ex}).  $\hat\mathfrak{C}_{\alpha J}$ itself is a sum, so 
for notational convenience define
\beqa 
\hat{\frak C}_{\alpha J;0\frac{1}{2}}&=
\sqrt{2} \sum_{\nu_2\nu_3}
D^{(1,1)}_{(0\nu_2\nu_3) \frac{1}{2};\alpha J}(\Omega^{-1})  \wz_{(0\nu_2\nu_3) \frac{1}{2}} \nonumber \\
&\qquad\qquad \times \left( \hat \ac^R_{ 0 \frac{1}{2}}(\lambda;\hat{\mathcal{C}}_2)
 +\hat \ac^L_{ 0 \frac{1}{2}}(\lambda;\hat{\mathcal{C}}_2)\right) \hat{\mathcal{C}}_2 ^{-1/2} \, ,\\
\hat{\frak C}_{\alpha J;2\frac{1}{2}}  &=
-\sqrt{2} \sum_{\nu_2\nu_3}
D^{(1,1)}_{(2\nu_2\nu_3) \frac{1}{2};\alpha J}(\Omega^{-1})  \wz_{(2\nu_2\nu_3) \frac{1}{2}} \nonumber \\
&\qquad\qquad \left( \hat \ac^L_{ 2 \frac{1}{2}}(\lambda;\hat{\mathcal{C}}_2) 
+\hat \ac^R_{ 2 \frac{1}{2}}(\lambda;\hat{\mathcal{C}}_2)  \right)
\hat{\mathcal{C}}_2 ^{-1/2}\, ,\\
\hat{\frak C}_{\alpha J;11}  &=
 \sum_{\nu_2 \nu_3} D^{(1,1)}_{(1\nu_2\nu_3) 1;\alpha J}(\Omega^{-1})
\sqrt{\frac{6}{(1+\delta_{\nu_2 1}\delta_{\nu_3 1})}}
\wh_{1\nu_2\nu_3;1} \nonumber \\
&\qquad\qquad\times 
\left( \hat \ac^L_{1 1}(\lambda;\hat{\mathcal{C}}_2)  +\hat \ac^R_{1 1}(\lambda;\hat{\mathcal{C}}_2)
\right)\hat{\mathcal{C}}_2 ^{-1} \, ,\\
\hat{\frak C}_{\alpha J;10} 
&= D^{(1,1)}_{(111)0;\alpha J}(\Omega^{-1})
\left( \hat \ac^L_{1 0}(\lambda;\hat{\mathcal{C}}_2)  +\hat \ac^R_{1 0}(\lambda;\hat{\mathcal{C}}_2)
\right) \, ,\\
\hat{\frak C}_{\alpha J}&=\hat{\frak C}_{\alpha J;0\frac{1}{2}}
+\hat{\frak C}_{\alpha J;2\frac{1}{2}}+\hat{\frak C}_{\alpha J;11}+
\hat{\frak C}_{\alpha J;10}
\eeqa 

Note that 
\beqa 
&&\left(\hat \ac^L_{0\frac{1}{2}}(\lambda;\hat{\cal C}_2) +\hat \ac^R_{ 0\frac{1}{2}}(\lambda;\hat{\cal C}_2) \right)
\hat{\cal C}_2^{-1/2} \, g(\Omega) \nonumber \\
&&\quad =
\left(\hat \ac^L_{2\frac{1}{2}}(\lambda;\hat{\cal C}_2) +\hat \ac^R_{ 2 \frac{1}{2}}(\lambda;\hat{\cal C}_2) \right)
\hat{\cal C}_2^{-1/2} \, g(\Omega)
\eeqa 
acting on any $g(\Omega)$, so that
\beqa 
&&
\left(\hat\frak{C}_{\alpha J;0\frac{1}{2}}+\hat\frak{C}_{\alpha J;2\frac{1}{2}}
\right)g(\Omega)\nonumber \\
&&= \sqrt{2} \sum_{\nu_2\nu_3}
\left(
D^{(1,1)}_{(0\nu_2\nu_3) \frac{1}{2};\alpha J}(\Omega^{-1})  \wz_{(0\nu_2\nu_3) \frac{1}{2}}
-D^{(1,1)}_{(2\nu_2\nu_3) \frac{1}{2};\alpha J}(\Omega^{-1})  \wz_{(2\nu_2\nu_3) \frac{1}{2}}
\right)\nonumber \\
& &\quad\times \left(\hat \ac^L_{0\frac{1}{2}}(\lambda;\hat{\cal C}_2) +\hat \ac^R_{ 0\frac{1}{2}}(\lambda;\hat{\cal C}_2) \right)
\hat{\cal C}_2^{-1/2}\,g(\Omega) \,
\eeqa

\subsection{$(\alpha J)=(111)0$.}  

This is the case where $\hat H=(T^{\lambda}_{(111)0})^2$ in Eq.(\ref{H2ex}).  
Let $g(\Omega)=g(\alpha_1,\beta_1,\alpha_2,\beta_2)$ be otherwise 
arbitrary; then we have
\beq 
\hat {\cal S}_{(111)0} \,\hat{\frak C}_{(111)0;\nu_1 I}\, g(\Omega) 
=  i\sqrt{ \frac{3}{2}}\frac{\partial }{\partial \alpha_2}\,\hat{\frak C}_{(111)0;\nu_1 I}\,
g(\Omega) 
\eeq 
Thus, we concentrate on the action of $\hat{\frak C}_{(111)0;\nu_1 I}$ on 
$g(\Omega)$.  For the $(\nu',I)=(0,\frac{1}{2})$ and $(2,\frac{1}{2})$ we obtain
\beq
\left(\sum_{\nu^\prime_1=0,2}\hat{\frak C}_{(111)0;\nu^\prime_1\frac{1}{2}}\right)\, 
g(\Omega)   =-\sqrt{3}\sin\beta_2\frac{\partial }{\partial \beta_2}\,
\left(\hat{\cal C}_2^{-1/2} \, g(\Omega) \right)
\eeq

As to the terms with $(\nu_1^\prime,I)=(1,1)$ and $(1,0)$, we find
\beqa 
&&\hat{\frak C}_{(111)0;11}g(\Omega) \nonumber \\
&&=\left(c^{(111)0}_{\beta_2}\frac{\partial}{\partial \beta_2}
+c^{(111)0}_{\beta_1} \frac{\partial}{\partial \beta_1} 
+c^{(111)0}_{(\alpha_2,\alpha_2)}
\frac{\partial^2}{\partial \alpha_2^2} 
+c^{(111)0}_{(\alpha_1,\alpha_2)}\frac{\partial^2}{\partial\alpha_1\partial\alpha_2}\right. \nonumber \\
&&\qquad\qquad\qquad  \left. +c^{(111)0}_{(\beta_2,\beta_2)}\frac{\partial^2}{\partial \beta_2^2} +
c^{(111)0}_{(\beta_1,\beta_1)}
\frac{\partial^2}{\partial \beta_1^2} +c^{(111)0}_{(\alpha_1,\alpha_1)} 
\frac{\partial^2}{\partial \alpha_1^2}\right) \nonumber \\
&&\qquad\qquad\qquad   \times \hat{\cal C}_2^{-1}
\left(\hat \ac^L_{11}(\lambda;\hat{\cal C}_2) +\hat \ac^R_{11}(\lambda;\hat{\cal C}_2) \right) 
g(\Omega)  \, ,
\label{c111011op}\\
&&\hat{\frak C}_{(111)0;10}g(\Omega) =
\frac{1}{4}(1+3\cos\beta_2)
\left( \hat \ac^L_{1 0}(\lambda;\hat{\mathcal{C}}_2)  +\hat \ac^R_{1 0}(\lambda;\hat{\mathcal{C}}_2)
\right)g(\Omega) \, ,
\eeqa 
where the coefficients $c^{(111)0}_{\Omega_k}$ and $c^{(111)0}_{(\Omega_j,\Omega_k)}$ are given in Table \ref{c111011coefficients}.

\begin{table}[h!]
\centering
{\renewcommand{\arraystretch}{2.5}
\begin{tabular}{|C|C||C|C|} \hline 
\Omega_i \hbox{ or } (\Omega_i,\Omega_j)& c_{\Omega_i}\hbox{ or } c_{\Omega_i\Omega_j}
&\Omega_i \hbox{ or } (\Omega_i,\Omega_j)& c_{\Omega_i}\hbox{ or } c_{\Omega_i\Omega_j}\\ 
\hline 
\beta_1 & \frac{3}{4}\tan\left(\textstyle\frac{1}{2}\beta_2\right)&
\beta_2 & \frac{3}{2}\cot\beta_1  \\
\hline 
(\alpha_2,\alpha_2)& 
\displaystyle
{\textstyle\frac{3}{8}}
\left(\tan^2({\textstyle\frac{1}{2}}\beta_1)-\frac{2}{1+\cos\beta_2}\right) & 
(\alpha_1,\alpha_2)&-\displaystyle\frac{3}{2(1+\cos\beta_1)} \\
\hline 
(\beta_2,\beta_2)&\frac{3}{4}(\cos\beta_2 -1) &
(\beta_1,\beta_1)&\frac{3}{2} \\
\hline 
(\alpha_1,\alpha_1)&\frac{3}{2}\csc^2(\beta_1) && \\ 
\hline
\end{tabular}
}
\caption{The non-zero function coefficients $c_{(\Omega_i\Omega_j)}$ 
in the operator $\frak{S}_{(111)0;11}=\sum_{ij}
c_{(\Omega_i\Omega_j)}{\partial^2}/{\partial\Omega_i\partial\Omega_j}$ of Eq.(\ref{c111011op}).}
\label{c111011coefficients}
\end{table}

\subsection{$(\alpha J)=(111)1$}  

This is the case where $\hat H=(\hat T^{\lambda}_{(111)1})^2$ in Eq.(\ref{H2ex}). 
Here, we have
\beq 
\left(\hat {\cal S}_{(111)1}\right)\,g(\Omega)
=\frac{i}{\sqrt{2}}\left( \frac{\partial }{\partial \alpha_2}
-2 \frac{\partial}{\partial \alpha_1}\right)
g(\Omega)
\eeq 

For terms with $\nu_1^\prime=0$ and $\nu'=2$:
\beqa 
&&\hat{\frak C}_{(111)1;0\frac{1}{2}}\,g(\Omega) \nonumber \\
&&\quad =
\left(-\frac{1}{2}\cos\beta_1\sin\beta_2\frac{\partial}{\partial \beta_2}
+\sin\beta_1\frac{\partial}{\partial \beta_1}
+\frac{i}{2}
\left( \frac{\partial }{\partial \alpha_2}
-2 \frac{\partial}{\partial \alpha_1}\right)
\right)\nonumber \\
&&\qquad\quad \times 
\left(\hat \ac^L_{0\frac{1}{2}}(\lambda;\hat {\cal C}_2) +\hat \ac^R_{ 0 \frac{1}{2}}(\lambda;\hat {\cal C}_2) \right)
\hat{\cal C}_2^{-1/2}g(\Omega)
\, ,\label{J1nuprime0} \\
&&\hat{\frak C}_{(111)1;2\frac{1}{2}}\,g(\Omega) \nonumber \\
&&=
\left(-\frac{1}{2}\cos\beta_1\sin\beta_2\frac{\partial}{\partial \beta_2}
+\sin\beta_1\frac{\partial}{\partial \beta_1}
-\frac{i}{2}
\left( \frac{\partial }{\partial \alpha_2}
-2 \frac{\partial}{\partial \alpha_1}\right)
\right)\nonumber \\
&&\qquad\quad \times 
\left(\hat \ac^L_{0\frac{1}{2}}(\lambda;\hat {\cal C}_2) +\hat \ac^R_{ 0 \frac{1}{2}}(\lambda;\hat {\cal C}_2) \right)
\hat{\cal C}_2^{-1/2}g(\Omega)\, ,
\eeqa 
the sum simplifies this time to
\beqa 
&&\left(\sum_{\nu^\prime_1=0,2}\hat{\frak C}_{(111)0;\nu^\prime_1\frac{1}{2}}\right)\, g(\Omega) \nonumber \\
&&\, =
\left(-\cos\beta_1\sin\beta_2\frac{\partial }{\partial \beta_2}+2\sin\beta_1
\frac{\partial}{\partial \beta_1}\right)\,
\hat{\cal C}_2^{-1/2} \, g(\Omega)
\eeqa

As to the terms with $\nu_1^\prime=1$, we find
\beqa 
&&\hat{\frak C}_{(111)1;11}g(\Omega) \nonumber \\
&&=\left[-\frac{\sqrt{3}\cos\left(\beta_1\right)
\left(3+\cos\left(\beta_2\right)\right)}{4\sin(\beta_2)}
\left(
\frac{\partial}{\partial \beta_2}+2\cot\left(\beta_1\right)\cot\left(
\textstyle\frac{1}{2}\beta_2\right)\frac{\partial }{\partial \beta_1}\right)
\right. \nonumber \\
&&+\frac{\sqrt{3}}{8}\left[2 \csc ^2\left(\textstyle\frac{1}{2}{\beta _2}\right) 
\sec ^2\left(\textstyle\frac{1}{2}\beta _1\right)+\cos \left(\beta _1\right) 
\left(\tan ^2\left(\textstyle\frac{1}{2}{\beta_1}\right)+\sec ^2\left(\textstyle\frac{1}{2}{\beta _2}\right)\right)\right]
\frac{\partial^2}{\partial \alpha_2^2} \nonumber \\
&& +
\frac{\sqrt{3}}{8}\left(\cos \left(\beta _1\right) \left(\cos \left(\beta _2\right)-1\right)-4\right) 
\csc ^2\left(\frac{\beta _2}{2}\right) \sec ^2\left(\frac{\beta_1}{2}\right)\frac{\partial^2}{\partial\alpha_1\partial\alpha_2}
\nonumber \\
&& \left. +\frac{3}{4}(\cos\beta_2 -1)\frac{\partial^2}{\partial \beta_2^2} 
-\frac{\sqrt{3}}{4}\cos \left(\beta _1\right) \left(\cos \left(\beta _2\right)+3\right) \csc ^2\left(\frac{\beta _2}{2}\right)
\frac{\partial^2}{\partial \beta_1^2} \right.
\nonumber \\
&& \left. +\frac{3}{2}\csc^2(\beta_1) \frac{\partial^2}{\partial \alpha_1^2}\right]  \hat{\cal C}_2^{-1}
\left(\hat \ac^L_{11}(\lambda;\hat{\cal C}_2) +\hat \ac^R_{11}(\lambda;\hat{\cal C}_2) \right) 
g(\Omega) \, ,\\
&&\hat{\frak C}_{(111)0;10}g(\Omega)\nonumber \\
&&\; =
\frac{1}{4}(1+3\cos\beta_2)
\left( \hat \ac^L_{1 0}(\lambda;\hat{\mathcal{C}}_2)  +\hat \ac^R_{1 0}(\lambda;\hat{\mathcal{C}}_2)
\right)g(\Omega)
\eeqa 



\bigskip

\begin{thebibliography}{99}


\bibitem{revPS} C.K. Zachos , D.B. Fairle and T.L. Curtright 2005 \textit{
Quantum mechanics in phase space} (World Scientific); A.M. Osorio de Almeida
Phys.Rep. \textbf{295} 265 (1998); Schroeck F 1996 Quantum Mechanics on Phase Space (Dordrecht: Kluwer)
\bibitem{suNKdeG}A.B. Klimov and H. de Guise, J.Phys. A 43, 402001 (2010).
\bibitem{WF} Wigner EP 1932 Phys. Rev. 40 749; Stratonovich RL 1956 Sov. Phys. JETP 31 1012; 
Hillery M, O'Connell RF, Scully MO and Wigner EP 1984 Phys. Rep. 106 121; 
Lee H-W 1995 Phys. Rep. 259 147;  
M. Gadella M 1995 Fortschr. Phys. 43, 229; 
G.S. Agarwal, Phys. Rev. A \textbf{24} 2889 (1981);
J.P. Dowling, G.S. Agarwal and W.P. Schleich, Phys. Rev. A 49, 4101 (1994).
\bibitem{brif} C. Brif and A. Mann , Phys. Rev .A \textbf{59} 971 (1999); Chaturvedi S, Ercolessi E, Marmo G, Morandi G, Mucunda N and Simon R (2006) J. Phys. A:
Math. Gen. 39 1405; Mucunda N, Marmo G, Zampini A, Chaturvedi S and Simon R (2005) J. Math. Phys. 46 012106.
\bibitem{Polkovnikov} Polkovnikov, Annals of Physics 325.8 (2010): 1790-1852; Davidson, Shainen M., and Anatoli Polkovnikov, Phys. Rev. Lett. 114.4 (2015): 045701;
Rundle, R. P., et al., Physical Review A 99.1 (2019): 012115; 
\bibitem{Tilma} T. Tilma, M. J. Everitt, J. H. Samson, W. J. Munro, and K.
Nemoto, Phys. Rev. Lett. \textbf{117}, 180401 (2016).
\bibitem{Onofri} Onofri, Enrico, Journal of Mathematical Physics 16.5 (1975): 1087-1089;
 F.T. Arecchi, E. Courtens, R. Gilmore and H. Thomas, Phys. Rev. A \textbf{6} 2211 (1972);
 \bibitem{coherentstatestuff} Gazeau, Jean-Pierre. Coherent states in quantum physics. Wiley, 2009;
Perelomov, Askold. Generalized coherent states and their applications. Springer Science \& Business Media, 2012;
Zhang, Wei-Min, and Robert Gilmore, Reviews of Modern Physics 62.4 (1990): 867;
\bibitem{book} A. B. Klimov, S.M. Chumakov, \emph{A Group-TheoreticalApproach to Quantum Optics},WILEY-VCH Verlag, Weinheimen, 2009.
\bibitem{review} A. B Klimov, J.L. Romero, H. de Guise 2017 J. Phys. A: Math. Theor. 50 323001
\bibitem{CGAlex} A.~C.~Nunes Martins, M.~W. Suffak and H. de~Guise, 
\emph{$SU(3)$ Clebsch-Gordan coefficients and some of their symmetries}, in preparation,
\bibitem{Moyal} J.E. Moyal, Proc. Camb. Phil. Soc. \textbf{45,} 99 (1949);Bayen F, Flato M, Fronsdal C, Lichnerowicz A and Sternheimer D 1978 Ann. Phys. (N.Y.) 111 61; 
Cattaneo A S, Felder G and Tomassini L (2002) Duke Math. J. 115 329?52
\bibitem{Belchev} Shirokov, Yu M, Fizika Ehlementarnykh Chastits i Atomnogo Yadra 10.1 (1979): 5-50;
Belchev, B., and M. A. Walton, 
Annals of Physics 324.3 (2009): 670-681;
Robbins, Matthew PG, and Mark A. Walton, Journal of Physics Communications 2.12 (2018): 125002.
\bibitem{su2} 
R. Gilmore, C.M. Bowden and L.M. Narducci, Phys.Rev. A \textbf{12} 1019 (1975); 
\bibitem{TWA} L.E. Ballentine, Y. Yang and J.P. Zibin Phys.Rev.A
\textbf{50} 2854 (1994); E.J. Heller Chem.Phys.\textbf{65} 1289 (1976); E.J.Heller
Chem.Phys.\textbf{67} 3339 (1977); E.J. Heller, J.R. Remiers and J.
Drolshangen Phys.Rev.A \textbf{190} 2613 (1987); Davis M J and Heller E J,
J. Chem. Phys. \textbf{80} 5036 (1984);  J.P. Amiet and M.B.
Cibils, J. Phys. A \textbf{24} (1991); P. Kinsler and P.D. Drummond
Phys.Rev.A \textbf{48} 3310 (1993);  G. Drobny and I. Jex, Phys.Rev A. \textbf{%
46} 499 (1992); G. Drobny, A. Bandilla and I. Jex, Phys.Rev.A\textbf{45} 78
(1996); A.B. Klimov, P. Espinoza, J. Opt. B: Quant.
Semiclass. Opt. \textbf{7} 183, (2005); A. Polkovnikov, Ann. Phys. \textbf{325} 1790 (2010); I.F. Valtierra, J.-L. Romero, A.B. Klimov, Phys.Rev. A 94 042336 (2016).
\bibitem{klimov} A.B. Klimov, J. Math. Phys. \textbf{43} 2202 (2002); A.B.
Klimov and P. Espinoza, J. Phys. A \textbf{35} 8435 (2002); P.M. Rios, and
E. Straume, \textit{Symbol Correspondences for Spin Systems}, Birkhauser (
Springer, International Publishing, Switzerland) (2014); B. Koczor, R. Zeier, S. J. Glaser, J. Phys. A: Math. Theor. 52 055302 (2019).
\bibitem{boop} Bopp F 1956 Ann. Inst. H. Poincare 15.
\bibitem{stratonovich} Stratonovich R L 1956 Sov. Phys.JETP 31 1012
\bibitem{bondia} J.C. V\'{a}rilly and J.M. Gardia-Bond\'{\i}a, Ann. Phys.\textbf{190} 107 (1989).
 \bibitem{su2CR} D. Zueco and I. Calvo, J. Phys. A\textbf{40} 4635 (2007); 
\bibitem{suN} Luis A 2008 J. Phys. A 41 495302; C. Braun and A. Garg, J. Math. Phys., \textbf{48} 032104
(2007); E. A. Kochetov. J. Math. Phys., \textbf{36} 1666 (1995); E. A.
Kochetov. J. Phys. A: Math. Gen., \textbf{31} 4473 (1998); T. F. Viscondi,
A. Grigolo, and M. A. M. de Aguiar J. Chem. Phys. \textbf{144} 094106 (2016); T.Tilma and K. Nemoto 2012 J. Phys. A: Math. Theor. 45 015302.
\bibitem{su3evol} A.B. Klimov, H.T. Dinani, Z.E.D. Medendorp and H.
de Guise, New. J. Phys \textbf{13} 113033 (2012); 
\bibitem{factorization} Rowe, D. J., B. C. Sanders, and H. de Guise, Journal of Mathematical Physics 40.7 (1999): 3604-3615. 
\bibitem{generalfactorization} de Guise, Hubert, Olivia Di Matteo, and Luis L. S\'{a}nchez-Soto,Physical Review A 97.2 (2018): 022328.
\bibitem{su3decomposition} Michael F O'Reilly Journal of Mathematical Physics, 23(11):2022-2028, 1982;
D Speiser. Theory of compact Lie groups and some applications to elementary particle physics.
Gordon and Breach, 1964; Maria SM Wesslen Journal of Mathematical Physics, 49(7):073506, 2008.

\bibitem{su3SS} Klimov, AB, Dinani, HT and de Guise, H, 2013, J.Phys.A, 46105302.

\bibitem{Varshalovich} Varshalovich, Dmitrii­ Aleksandrovich, Anatolij Nikolaevii Moskalev, and Valerii Kel'manovich Khersonskii. Quantum theory of angular momentum. 1988.

\bibitem{hdg2011} Klimov, Andrei B., et al, New Journal of Physics 13.11 (2011): 113033.

\bibitem{DraayerAkiyama} Draayer, J. P., and Yoshimi Akiyama, Journal of Mathematical Physics 14.12 (1973): 1904-1912.

\bibitem{BEstuff} T.F. Viscondi, K. Furuya and M. C. de Oliveira, European Physics Letters, 90 (2010) 10014;
Roberto Franzosi and Vittoria Penna, Physical Review A 65 (2001): 013601; K. Nemoto, C. A. Holmes, G. J. Milburn and
W. J. Munro, Physical Review A 63 (2000) 013604; S. Mossmann and C. Jung, Physical Review A 74 (2006) 033601.
 
\end{thebibliography}
\end{document}